\documentclass[fleqn,usenatbib]{mnras}

\usepackage{newtxtext,newtxmath}
\usepackage{xcolor}
\usepackage[T1]{fontenc}

\usepackage{longtable}

\DeclareRobustCommand{\VAN}[3]{#2}
\let\VANthebibliography\thebibliography
\def\thebibliography{\DeclareRobustCommand{\VAN}[3]{##3}\VANthebibliography}

\usepackage{graphicx}	
\usepackage{amsmath}	

\def\splus{S-PLUS}

\title[STEP Survey]{The S-PLUS Transient Extension Program: Imaging Pipeline, Transient Identification, and Survey Optimization for Multi-Messenger Astronomy}

\author[A. Santos et al.]{
A. Santos$^{1}$,\thanks{E-mail: andsantos@cbpf.br}
C. D. Kilpatrick$^{2}$,
C. R. Bom$^{1,3}$,
P. Darc$^{1}$,
F. R. Herpich$^{10}$,
E. A. D. Lacerda$^{4}$,
M. J. Sartori$^{4}$,
\newauthor
A. Alvarez-Candal$^{5,6}$,
C. Mendes de Oliveira$^{4}$,
A. Kanaan$^{7}$, 
T. Ribeiro$^{8}$,
W. Schoenell$^{9}$\\
$^{1}$Centro Brasileiro de Pesquisas Físicas, Rua Dr. Xavier Sigaud 150, CEP 22290-180, Rio de Janeiro, RJ, Brazil\\
$^{2}$Center for Interdisciplinary Exploration and Research in Astrophysics (CIERA) and Department of Physics and Astronomy,\\
Northwestern University, Evanston, IL 60208, USA\\
$^{3}$Centro Federal de Educa\c{c}\~ao Tecnol\'ogica Celso Suckow da Fonseca, Rodovia M\'ario Covas, lote J2, quadra J, CEP 23810-000,  Itagua\'i, RJ, Brazil\\
$^{4}$Universidade de S\~ao Paulo, IAG, Rua do Mato 1225, Sao
Paulo, SP, Brazil\\
$^{5}$Instituto de Astrof\'isica de Andaluc\'ia, CSIC, Apt 3004, E18080 Granada, Spain\\
$^{6}$Instituto de F\'isica Aplicada a las Ciencias y las Tecnolog\'ias, Universidad de Alicante, San Vicent del Raspeig, E03080, Alicante, Spain\\
$^{7}$Departamento de F\'isica, Universidade Federal de Santa Catarina, Florian\'opolis, SC, 88040-900, Brazil \\
$^{8}$ Departamento de Astronomia, Instituto de F\'isica, Universidade Federal do Rio Grande do Sul (UFRGS), Av. Bento Goncalves 9500, Porto Alegre, RS, Brazil\\
$^{9}$NOAO, P.O. Box 26732, Tucson, AZ 85726, USA\\
$^{10}$Cambridge Survey Astronomical Unit (CASU), Institute of Astronomy, Madingley Road, Cambridge CB3 0HA, UK
}

\date{Accepted XXX. Received YYY; in original form ZZZ}

\pubyear{2023}

\begin{document}
\label{firstpage}
\pagerange{\pageref{firstpage}--\pageref{lastpage}}
\maketitle

\begin{abstract}

We present the S-PLUS Transient Extension Program (STEP): a supernova and fast transient survey conducted in the southern hemisphere using data from the Southern Photometric Local Universe Survey (S-PLUS) Main Survey and the T80-South telescope. Transient astrophysical phenomena have a range of interest that goes through different fields of astrophysics and cosmology. With the detection of an electromagnetic counterpart to the gravitational wave (GW) event GW170817 from a binary neutron stars merger, new techniques and resources to study fast astrophysical transients in the multi-messenger context have increased. In this paper, we present the STEP overview, the SN follow-up data obtained, data reduction, analysis of new transients and deep learning algorithms to optimize transient candidate selection. Additionally, we present prospects and optimized strategy for the search of Gravitational Wave counterparts in the current LIGO/Virgo/Kagra observational run (O4) in the context of T80-South telescope.

\end{abstract}

\begin{keywords}
surveys -- transients: supernovae -- gravitational waves
\end{keywords}


\section{Introduction}\label{sec:introduction}

Transient surveys have become increasingly important in astronomy, astrophysics and cosmology due to their ability to provide valuable information on astrophysical time-dependent phenomena such as supernovae, variable stars, active galactic nuclei (AGN), tidal disruption events, gamma-ray bursts, and kilonovae \citep[e.g.][]{andreoni2022very,stein_23,YSNE,yang_kn23}.  These events span a wide range of luminosities, spectral types, and timescales, often requiring survey strategies uniquely tailored to find specific classes of transients. By systematically observing and analyzing these transient phenomena, astrophysicists can unravel the dynamic process of those events and gain a deeper understanding of their intricacies.

As transient surveys have advanced, the development of technologies such as charge-coupled devices (CCDs) and robotic telescopes has greatly facilitated the systematic search for astrophysical transients \citep{Perlmutter89, Filippenko92, Richmond93}. Historically, one of the earlier studies about a systematic approach for searching extragalactic sources was developed by \cite{zwicky1938search}, increasing the number of discovered events over the next decades. Searches in the southern hemisphere started around 1980 by \citep{1980tsup.work....7M}. These advancements enable more efficient and automated detection of transient events, significantly increasing the discovery rate, and allowing detailed follow-up observations and analysis. With the combined power of wide-field observations, advanced instruments, and data analysis techniques, transient surveys push the boundaries in astrophysics research, unveiling new phenomena and providing a comprehensive view of the ever-evolving universe.

Additionally, the detection, classification, and analysis of Supernova type Ia changed our understanding of the Universe, by serving as "standard candles" \citep{branch1992type}. With consistent luminosity distances, Type Ia supernovae allow astronomers to accurately measure cosmic distances, which ultimately led to the groundbreaking discovery of the universe's accelerated expansion and therefore insights about a novel component known as dark energy \citep{1929PNAS...15..168H}. Transient surveys have been instrumental in discovering and characterizing these important cosmic explosions arising from massive stars, compact objects, and stellar interactions \citep[e.g.,][]{Hamuy93,Filippenko05,Kulkarni07,Bellm19,Karambelkar23}. Light curve data from those celestial objects plays a crucial role in tailoring fundamental aspects of explosive events, such as their energy levels and mass ejecta \citep{Woosley88,Gal-Yam09}, constraints of the presence of circumstellar matter in their environments \citep{Smith14}, and the formation of compact objects \citep{MacFadyen01,Soderberg10}. It's also possible to use explosive transients for cosmology cases \citep{Riess98, Perlmutter99}.
\par Beyond the supernovae, transient surveys also contribute to the study of gravitational waves (GWs). The detection of GW events, such as the groundbreaking GW170817, which was initially detected in GW emission \citep{Abbott17:mma} and precisely localized at optical wavelengths \citep{soares2017electromagnetic} opened new opportunities to study the compact objects that lead to LIGO/Virgo GW events. GWs provide unique insights into some of the most energetic and cataclysmic events, including their contribution to Galactic chemical enrichment \citep{Eichler89,Argast04,Arnould07,Metzger10,Kasen17} and formation of heavier neutron starts (NS) and black holes (BH) \citep{Belczynski02,Dominik15}. Furthermore, the use of gravitational waves in cosmology allows for precise measurements of the Hubble Constant \citep{2017Natur.551...85A,Soares-Santos_2019,chen2018two, bom23} and provides valuable constraints on cosmological models.
With the increased focus on optical transient surveys for studying the physics of explosive transients, multi-messenger astronomy and cosmology, individual surveys are increasingly tailored to focus on some subset of depth, cadence, total area covered and wavelength coverage in order to maximize their own science returns.  For example, the Vera C. Rubin's Legacy Survey of Space and Time \citep[LSST;][]{LSST09,LSST} will observe the Southern hemisphere with unprecedented depth for a wide-field optical time-domain survey, but without the cadence necessary to find and characterize rapid transients such as kilonovae \citep{Andreoni22} and saturating at $\approx$17~mag.  The synergy between LSST and other wide-field optical survey and follow-up telescopes, especially when their data streams are combined using the latest transient alert brokers and target and observation managers \citep[TOMs; e.g.,][]{Volgenau22,Coulter23}, will enable new science that either survey cannot perform alone.

Here we describe an optical time-domain survey using the T80-South telescope located at Cerro Tololo, using imaging from the main S-PLUS Survey, which we call the S-PLUS Transient Extension Program (STEP, PIs: Bom \& Kilpatrick). We describe our pipeline and image processing procedures for retrieving and processing image data as well as our methods for identifying and reporting transient candidates.  We then describe several science cases that highlight the utility of STEP, particularly for discovering supernovae, follow up and characterization based on their light curves, and follow up of GW events during Ligo-Virgo-KAGRA (LVK) observing runs.

The paper is structured as follows: in Section 2 we describe our scientific goals with this project and how STEP is designed to achieve those goals. In Section 3, we describe each step in our imaging and analysis pipeline starting from obtaining images from the telescope, our reduction process, our difference imaging analysis procedure and finally candidate selection. In Section 4 we detail our deep learning algorithm designed to reject false positives during difference image analysis. In Section 5 we outline extragalactic transients detected during our project. In section 6 we overview the light curve cadence of followed SNe. In section 7 we talk about our strategy and criteria to follow GW events. Finally, in Section 8 we review our conclusions and describe our future plans for this project in the era of LSST time-domain discovery and follow up.
\section{STEP Science Goals}\label{sec:step_science_goals}

\subsection{\splus\ Main Survey Overview and Discovery of Transients in the Southern Hemisphere}
The Southern Photometric Local Universe Survey (\splus; \citealt{MendesDeOliveira+19}) consists of an imaging survey that will cover approximately $9300$ deg$^{2}$ in 12 optical bands ($ugriz$ and seven narrow filters). The survey utilizes a dedicated 0.8-m aperture robotic telescope called T80-South (T80S) located at Cerro Tololo Interamerican Observatory (CTIO), Chile. The sky coverage is divided into five sub-surveys, two of which are relevant to this project: The Main Survey (MS), and the Galactic Survey (GS).
The MS covers 8300~deg$^{2}$ and follows an observing strategy tailored for extragalactic science, providing photometric redshifts down to $i$=21~mag with a limiting magnitude up to 21.5~mag in $r$-band \citep{AlmeidaFernandes+22}. The Galactic Survey (GS) spans an area of 1300~deg$^{2}$ in the Milky Way Plane in all 12 filters, including two galactic regions - namely the bulge and the disk - and employs the same observational strategy as MS. T80S have an e2v detector with a 9232 $\times$ 9216 array with 10 $\rm \mu m$ pixel, a plate scale of  0.55\arcsec~pixel$^{-1}$ and a field-of-view (FoV) of $1.4 \times 1.4$~$ \mbox{deg}^{2}$.   In total, during our first year of operation, STEP analyzed 1272~deg$^{2}$ in $griz$ to a median depth of $\approx$20.5~mag. We outline the light curve of three followed SNe in appendix \ref{appendix:photometry}, as our first data release (DR1).

The MS employs a single epoch observation of each field per filter, with a range of seeing from 0.8 to 2 arcsec. In this survey, three exposures are taken for each filter with an exposure time that goes between $\sim30$ and $\sim60$ seconds for the \textit{griz} broad bands (see Table 5 of \citealt{MendesDeOliveira+19} for exact exposure time of all T80cam filters), dithering a few arcseconds to the east and west of the field center. The primary objective of this strategy is to enhance the accuracy of photometric redshift estimation for objects within the field, targeting those with absolute magnitudes down to $r_{AB} \sim20$. Furthermore, the redshift information provided by this approach aims to achieve a precision of $\Delta z/(1+z) = 0.02$. The individually captured images are overscan-subtracted, trimmed, bias-subtracted, flat-divided, and fringing subtracted ($z$-band only; see \citealt{MendesDeOliveira+19} for more details on the reduction process and pipeline). Besides these steps, the images are cosmetic-corrected to identify and mask satellite tracks and cosmic rays and then astrometrically solved(see \citealt{AlmeidaFernandes+22} for more on the astrometric precision).

Given the previous description of S-PLUS and T80cam, its wide field-of-view camera, and an excellent calibration of its photometric system \citep{MendesDeOliveira+19}, \splus\ emerges as a suitable survey for investigating transient phenomena in the Southern sky. Moreover, the survey benefits from its overlap with other time-domain surveys including DES, ATLAS, and PAN-STARRS, which collectively cover various portions of the Southern sky. This survey also fills a critical gap in time-domain coverage of the Southern sky before the Vera C. Rubin Observatory \citep{LSST} and La Silla Schmidt Southern Survey (LS4) come online to cover this sky area with high cadence. Given all this information, we argue that \splus\ is an appropriate instrument for investigating transient phenomena in the Southern sky.

\subsection{Supernova Follow Up}

Supernova follow up in the era of the Rubin Observatory will rely on careful selection of transients from the thousands of potentially valuable transients that will be discovered each night. Even restricting those samples to events accessible by smaller aperture telescopes such as T80S requires selecting from potentially hundreds of events and following those necessary for specific science cases, such as measuring cosmological parameters from SN\,Ia light curves, constructing SN luminosity functions, testing light curve classifiers, or constraining the incidence of strong circumstellar interaction in core-collapse SNe.  All of these science cases can be addressed by synthesizing data from multiple sources using transient brokers \citep[e.g.,][]{Narayan18,Forster21,Moller21}, photometric classification of light curves \citep{Boone19,Muthukrishna19,Villar19,Sravan20}, and anomaly identification to prioritize transients with intrinsically peculiar properties (e.g., total ejecta mass or energy, nickel mass, incidence of circumstellar interaction) or those at critical or scientifically useful phases.

Obtaining multi-band light curves for the thousands of transients needed for light curve parameter estimation will require coordination between a global network of telescopes.  Telescopes such as T80S can improve the effective cadence of Rubin Observatory by observing between visits from the survey.  Moreover, large samples of transient light curves are needed {\it now} to train photometric classifiers and anomaly detection.  We present supernova follow up in Section~\ref{sec:supernova} and further discuss prospects for follow up in the Rubin era in Section~\ref{sec:conclusions}.

\subsection{Gravitational Wave Follow Up During LIGO/Virgo/KAGRA Observing Run 4}

The kilonova counterpart to GW170817 was rapidly declining and required rapid, high-cadence, multi-band follow up to constrain its total ejecta mass and composition, including observations from the T80S \citep[see, e.g.,][]{Abbott17:mma,Coulter17,Diaz17,Kilpatrick17,soares2017electromagnetic}. Due to the increased sensitivity of LIGO since that time, NS mergers found during observing run 4 (O4) are systematically more distant than GW170817 at 40\,Mpc \citep{Abbott17:mma,soares2017electromagnetic} and may in general be more poorly localized.  As these events may occur and be detected anywhere in the sky, a global network of robotic telescopes with wide fields of view and multi-band follow up are needed to localize and obtain optical follow up of these events.  Within this context, we consider the feasibility of future GW follow up to identify new kilonovae with the T80S.

Numerous dedicated efforts have been undertaken since the localization of the GW170817 kilonova to identify additional EM counterparts associated with similar NS mergers \citep[e.g., for events discovered during observing run 3;][]{Andreoni19,Andreoni19b,Coughlin19,Dobie19,Goldstein19b,Gomez19,Hosseinzadeh19,Lundquist19,Ackley20,Antier20,Coughlin20b,Morgan20,Paterson20,Pozanenko20,Thakur20,Vieira20,Watson20,Alexander21,deWet21,Kilpatrick21,Tucker21}. To improve the observing efficiency of these search efforts and aid in rapidly finding kilonova to maximize the scientific opportunities that are possible with their early light curves \citep[e.g.,][]{Arcavi18}, we consider a new observing strategy optimized for the sensitivity, field of view, and broadband filters on the T80S. Considering this strategy in combination with realistic selection criteria for O4 GW events, we consider the observations necessary to find the next kilonova counterpart in Section~\ref{sec:gw_follow_up}.
\section{STEP Pipeline Overview and Data Analysis}\label{sec:step_pipeline_overview}

The STEP pipeline uses two types of images: archive images and on-the-fly observations. Archive images were obtained before the STEP project started and found in their final reduction form (updated calibration frames, astrometry, and quality checked). These are the images co-added to compose the \splus\ Main Survey. The on-the-fly images are those run through the STEP pipeline as soon as they are observed to enable rapid identification of transients. They are usually ``last night images'' and are not yet the final form of the reduction process, using older bias and flat frames. They also do not include fringe corrections. Although not the final version, they enable STEP to access the most up-to-date picture of the observed sky and identify possible transient phenomena as they are ongoing, which allows us to follow up on the explosion to acquire further information about the phenomena.

\begin{figure*}
    \includegraphics[width=0.8\linewidth]{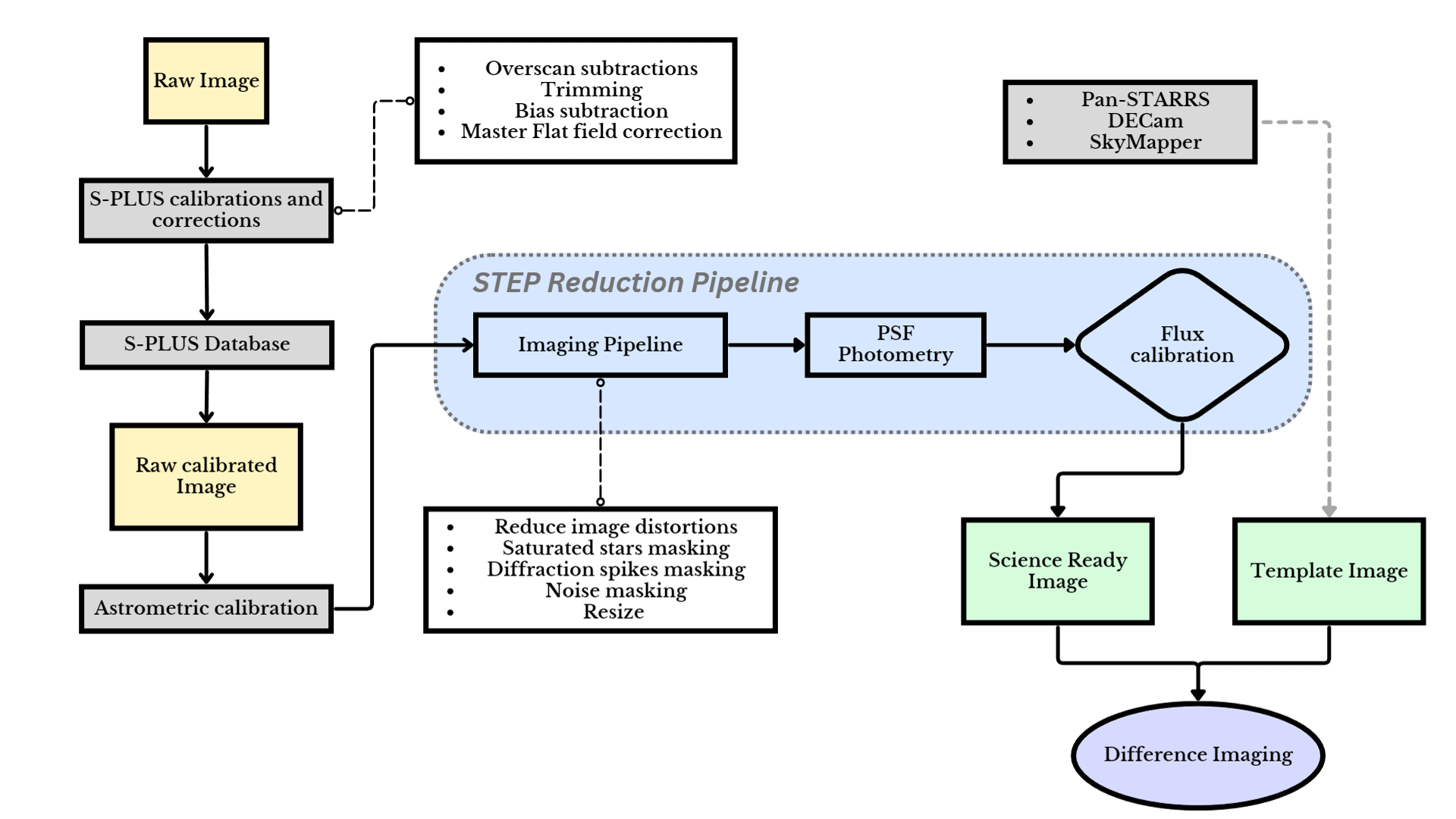}
    \centering
    \caption{A schematic overview of the reduction pipeline. The raw image from S-PLUS goes through pre-processing like overscan, bias and flat field correction, and satellite track removals. We retrieve those images and initiate our reduction pipeline for further analysis. Our pipeline goes through an astrometric solution of all retrieved images. We then proceed to masking of saturated stars, diffraction spikes, and sky background (noise). After the masking stages, We resized all images to a standard of 10.000 x 10.000 points. Finally, our images go through PSF photometry and Flux calibration. The template images go through similar stages (with the exception of diffraction spikes masking and astrometric solution because those steps are not needed in this case. In the end, we perform the difference imaging as a final image product.}
    \label{fig:workflow}
\end{figure*}

Our methodology to discover transient candidates involves utilizing the comprehensive coverage of the main survey fields in the \textit{griz} filters to extract information on potential transients through a state-of-the-art pipeline described below and summarized in Figure \ref{fig:workflow}. Our objective is to document and characterize transients identified within these images, leveraging the photometric redshift to derive the luminosity distance and other intrinsic properties of the object, as well as complementary spectroscopy and host galaxy data, when available. The \splus\ Main Survey is specifically designed to encompass a wide expanse of the southern hemisphere, reaching high declination angles. This expansive coverage facilitates the identification of transient phenomena in the region.

The STEP pipeline operates by processing nightly data from S-PLUS, taking advantage of its ongoing survey to look for transients. S-PLUS does not revisit the same field, thus to obtain template images, we use templates from other surveys that overlap with S-PLUS footprint (e.g. DECam, Pan-STARRS, and SkyMapper). We transfer and digitally subtract our collected images from S-PLUS with the templates derived from those other surveys. By employing this technique, we can identify transient sources within approximately 1 hour of image acquisition. Each identified source is then carefully evaluated using source properties within each image frame, contextual data extracted from the star, minor planet, and galaxy catalogs, and machine learning classification. We describe our use of the {\it Gaia} DR3 and minor planet checker in Section~\ref{sec:candidate-select}, while the galaxy catalogs used are those that provide context to the distance and thus luminosity scale of each transient detection via spectroscopic or photometric redshifts.  Specifically, we make use of the Pan-STARRS Source Types and Redshifts with Machine Learning (STRM) catalog \citep{Beck21}, the Legacy photometric redshift catalog \citep{Zhou20}, and the S-PLUS photometric redshift catalog \citep{MendesDeOliveira+19}. This comprehensive approach ensures the thorough vetting of each transient source. The ultimate goal of our pipeline is to optimize follow-up observations for multi-messenger astronomy, making use of the multi-band photometry available from the S-PLUS Main Survey. The subsequent sections in this paper provide detailed explanations of each step within our pipeline, starting from image acquisition and extending through the transient identification process.

\subsection{Reduction Processing of T80S and Template Imaging}

Before the reduction process and analysis, we begin with the daily collection of science images from the S-PLUS Main Survey (and galactic survey, when available), captured using the \textit{griz} broad-band filters. The retrieved images have undergone a pre-reduction process, which encompasses several crucial steps, including bias subtraction, flat field correction, overscan subtraction, and trimming. These steps are implemented to enhance the overall quality of the images. Subsequently, the images are transferred to our servers, where an initial pre-processing stage is conducted before they are fed into the reduction process. For astrometric calibration of the raw images collected, we utilize the command-line tools provided by {\tt astrometry.net} \citep{astrometry.net}.

Following these pre-processing steps, the main STEP pipeline takes charge of the subsequent stages, encompassing imaging, photometry, and data analysis. Throughout the forthcoming sections, we provide detailed descriptions of each stage of our software, elucidating their functionalities, and highlighting their integration with essential open-source software tools prevalent within the astronomical community.

Template images are constructed for each unique field and imaging band by communicating the observed fields to a separate server.  Our pipeline utilizes a daily {\tt cron} job that first downloads images saved as {\tt FITS} files \citep{FITS} from the T80S telescope database for the previous night and communicates the field center and imaging band continuously to the template server.  The template server then determines if the field has coverage in Pan-STARRS, DECam, or SkyMapper, and downloads all available imaging covering that field from the Mikulski Archive for Space Telescopes (MAST) for Pan-STARRS \citep{2002SPIE.4836..154K}, the NOIRLab archive for DECam, or the SkyMapper image cutout service\footnote{\url{https://skymapper.anu.edu.au/image-cutout/}} \citep{skymapperDR1}.  To make the photometric calibration and point-spread function (PSF) shape as consistent as possible in each template, our pipeline uses only a single survey to generate each template.  From the input images, we perform sky subtraction, normalization using their reported photometric calibration, and convolution to a uniform PSF shape using {\tt SWarp} \citep{2010ascl.soft10068B} to produce an output template image with the input field center and pixel scale of the T80S camera.  The final template image is then staged in our pipeline, and a final photometric catalog is generated in the image frame using {\tt DoPhot} \citep{1993PASP..105.1342S} and a zero point empirically estimated for the image by comparing to the Pan-STARRS or SkyMapper \textit{griz} photometric catalogs.

For the T80S science frames, the pipeline uses {\tt IRAF} to estimate a geometric distortion solution across the detector using Gaia DR3 astrometric standards and algorithms implemented in {\tt msccmatch} \citep{IRAF}.  The final distortion solution is saved in {\tt TNX} format in the image header to regrid each frame to a common pixel scale and field center as the template images at a later stage.

We then use {\tt source extractor} \citep[{\tt sextractor};][]{1996A&AS..117..393B} to measure the background levels and generate a catalog of saturated stars. From this catalog and output segmentation maps in {\tt sextractor}, we mask out saturated stars and the unique diffraction spike pattern for T80S based on the number of saturated pixels within each star. The output file from this stage is a unique mask image created with {\tt mana}\footnote{\url{https://www.cfht.hawaii.edu/Instruments/Elixir/Ohana/mana.html}}. We then use {\tt SWarp} \citep{2010ascl.soft10068B} to regrid the science and mask images, assuring that the science and template images have a common and uniform pixel scale of 0.55\arcsec\ across the image, a common field center, and a uniform size of 10000$\times$10000 per frame.  As with the template images, the final stages of our science image pipeline perform PSF photometry using {\tt DoPhot} and flux calibration by comparing to the Pan-STARRS DR2 and SkyMapper photometric catalogs. In general, we assume that the bandpasses from each template survey are identical to those from the S-PLUS survey, that is, that S-PLUS $griz$ bands are identical to PS1, DECam, and SkyMapper $griz$ bands both for calibration and image subtraction.  We are exploring updates to our calibration methods for each photometric catalog and source of template images to correct for slight differences, for example, due to mirror coatings, bandpasses, and detector quantum efficiency.  Given the calibration systematics achieved from the overall 12-filter S-PLUS system \citep{MendesDeOliveira+19}, we expect to be able to achieve similar residuals as cross-calibration from other optical surveys \citep[e.g.,][]{Finkbeiner16,Scolnic15}, with cross-calibration residuals of a few millimag.

\subsection{Difference Imaging}

After the reduction of both science and template images, we match the template to science images to perform digital image subtraction or difference imaging. This stage is fundamental to finding transients as it enables the automatic detection of significant point sources present in the science image but not in the template (or vice versa).  Our difference image procedure utilizes the calibration science and template images, their noise frames, and mask images to perform subtraction in {\tt hotpants} \citep{2015ascl.soft04004B}, an open-source tool designed for this purpose.  {\tt hotpants} uses common point sources to define a Gaussian convolution kernel for PSF matching between the science and template images then directly subtracts the convolved image from the other image. In general, the template image is deeper than the science image. It has a sharper PSF, red in particular for the PS1 and DECam imaging that makeup $>$95\% of our template imaging.  For the remaining $\sim$5\% of our template imaging that comes from the 1.35\,m SkyMapper telescope \citep{skymapperDR1}, the template image stack is often still deeper than our T80S imaging, but further optimization of the STEP pipeline is ongoing to explore the effect of image depth for these data.  In general, we set the convolution direction to convolve the science image. We also use our point source catalog from {\tt DoPhot} to define the locations of ``substamps'' in {\tt hotpants}, that is the regions of the image used to estimate the parameters of the convolution kernel. Finally, given the large field-of-view of the T80S camera, we fit the convolution kernel using a third-order polynomial to estimate the spatial variation of the kernel across the image and a second-order polynomial to subtraction background emission. All other parameters are set to the defaults in {\tt hotpants}.

After the image subtraction process, we check the quality of the resultant difference image for the average scatter in the fitted kernel. Using the fixed PSF parameters of the template image, we then search for significant point sources and perform photometry using a customized version of {\tt DoPhot} for this process. This catalog is used to generate cutout images of possible transients to be posted on a webpage for visual inspection and to feed a Neural Network that classifies a given cutoff into real transient or bad subtraction.

\subsection{Transient Candidate Selection}\label{sec:candidate-select}

We identify all candidate transients in our data in every image frame where {\tt DoPhot} reports a significant detection, clustering these detections for sources within 1\arcsec\ across separate frames of the same field.  We then perform forced photometry across all frames in the same field at the flux-weighted centroids of each ``cluster'' using {\tt DoPhot}.  The unforced and forced photometric data, individual centroid positions, incidence of masking within the PSF aperture, cutouts of the science, template, and difference images around the position of the transient are displayed for each candidate in an internal webpage used for visual inspection (not publicly available). (Figure~\ref{fig:websniff}). These data were initially visually scanned for real/bogus identification of transients versus subtraction artifacts and other sources of non-astrophysical emission and cosmic rays (flagged as ``BAD'' sources). Our visual flagging criteria for all other transient sources are stored to generate a training sample for the machine learning process described in Section~\ref{sec:machine-learning}.

From this list of transients, we further vet candidates using methods implemented in \citep{candidates}. We use minor planet catalogs to flag known asteroids within 10\arcsec\ of the source using the Modified Julian Date and coordinates of each detection (``ASTEROID'' sources) using the Minor Planet Checker\footnote{\url{https://minorplanetcenter.net}}. We also check for coincidence ($<$1\arcsec) of any source classified as a likely, unblended point source in the {\it Gaia} DR3 catalog to flag sources of stellar variability \citep[``VARSTAR'' sources; i.e., by crossmatching to the {\it Gaia} point-source catalog in][]{GaiaDR3}. Finally, we crossmatch all candidates to known transients in the Transient Name Server database \citep{Yaron12} to flag all ``CONFIRMED'' transients. We incorporate all objects and our manually applied flags into the machine learning process described below. The final reported transients are the result of this deep learning candidate selection.

\section{Deep Learning Candidate selection}\label{sec:machine-learning}

In practice, most transient alerts for a given observation night are false positives as the difference imaging process produces various artifacts. The most common example of an artifact we classify as a ``Bad Subtraction''. When there is a mismatch between the flux from the search and template images, for example, due to slight differences in filter transmission between T80S-Cam and PS1, DECam, or SkyMapper, or if there are variations between the PSFs not properly accounted for by {\tt hotpants}, the subtraction process results in under- and over-subtracted regions in the difference image. The most common bad subtractions are bright stars that appear in the difference image as positive or negative point sources. We acknowledge that this phenomenon is difficult to distinguish from instances of true stellar variability (i.e., ``VARSTAR'' sources), but in general we find that $\sim$1\% of stars in our images exhibit variability $>$0.1~mag in optical bands, which is at least two orders of magnitude more common than stars observed by optical time-domain surveys such as {\it Kepler} \citep{McQuillan12}. We infer that the vast majority of these  transient alerts are bad subtractions and describe a method for filtering these alerts below.

We are primarily interested in reducing the number of candidates that require visual inspection and automating the transient detection process, for which we adopt a convolutional neural network (CNN) to classify the images as transients or artifacts (i.e., non-transients). Applying CNNs to difference images is a common approach in modern time-domain surveys \citep{Duev19,Shandonay22} with multiple well-tested procedures in the literature.  We present a new method for systematically classifying sources that are optimized to our pipeline and data reduction procedures.

To address this classification problem, we made use of a family of CNN models known as {\tt Mobilenet} \citep{Howard17} and prepared the data following the prescription of \citep{bom2022}. This architecture is based on depth-wise separable convolutions. This architecture demonstrates superior speed and accuracy characteristics in our data set when compared to several different CNN-based models such as {\tt VGG} \citep{Simonyan14}, {\tt ResNet} \citep{He2016} and {\tt Inception} models \citep{Szegedy2015}.

\subsection{Training, Validation, and Test Data Set}

\begin{figure}
    \includegraphics[width=\linewidth]{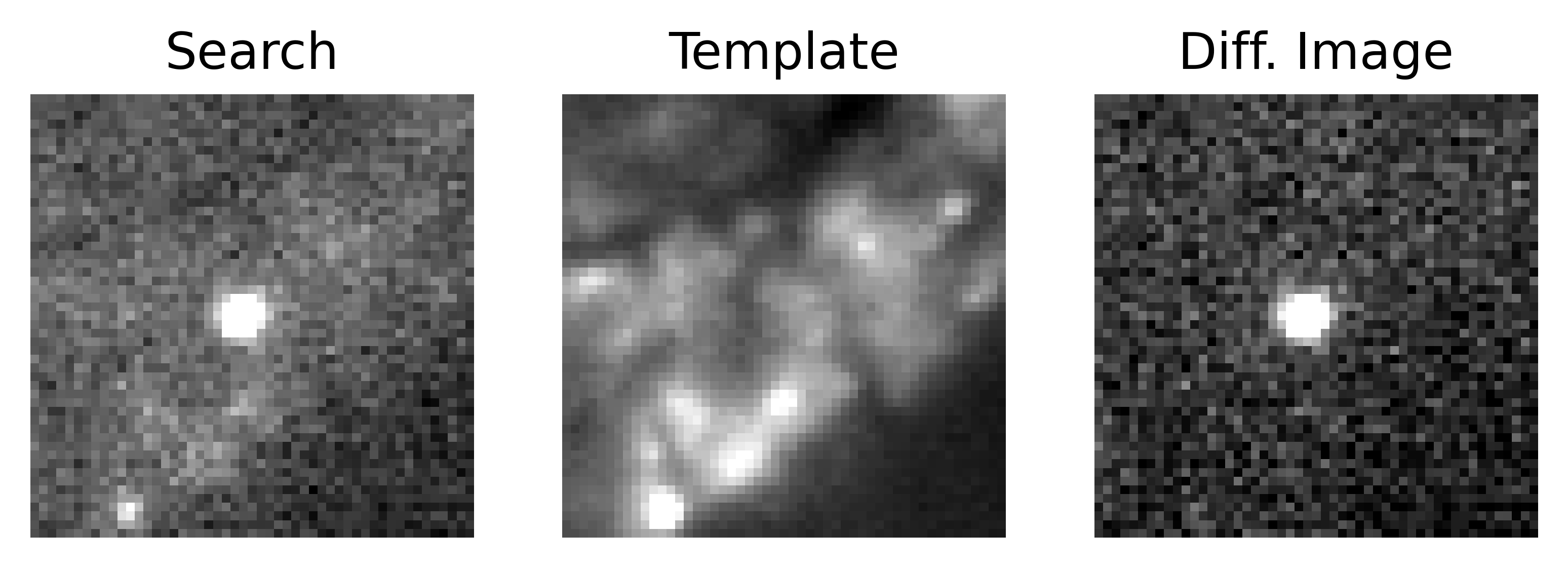}
    \includegraphics[width=\linewidth]{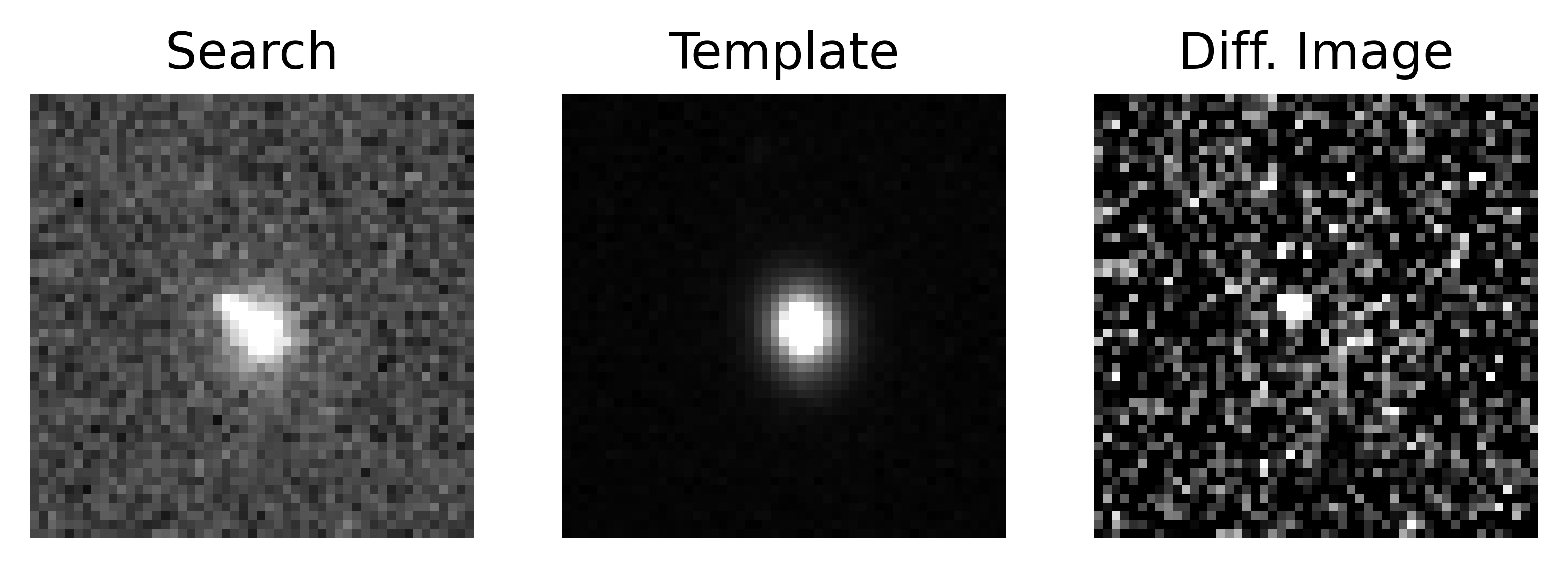}
    \includegraphics[width=\linewidth]{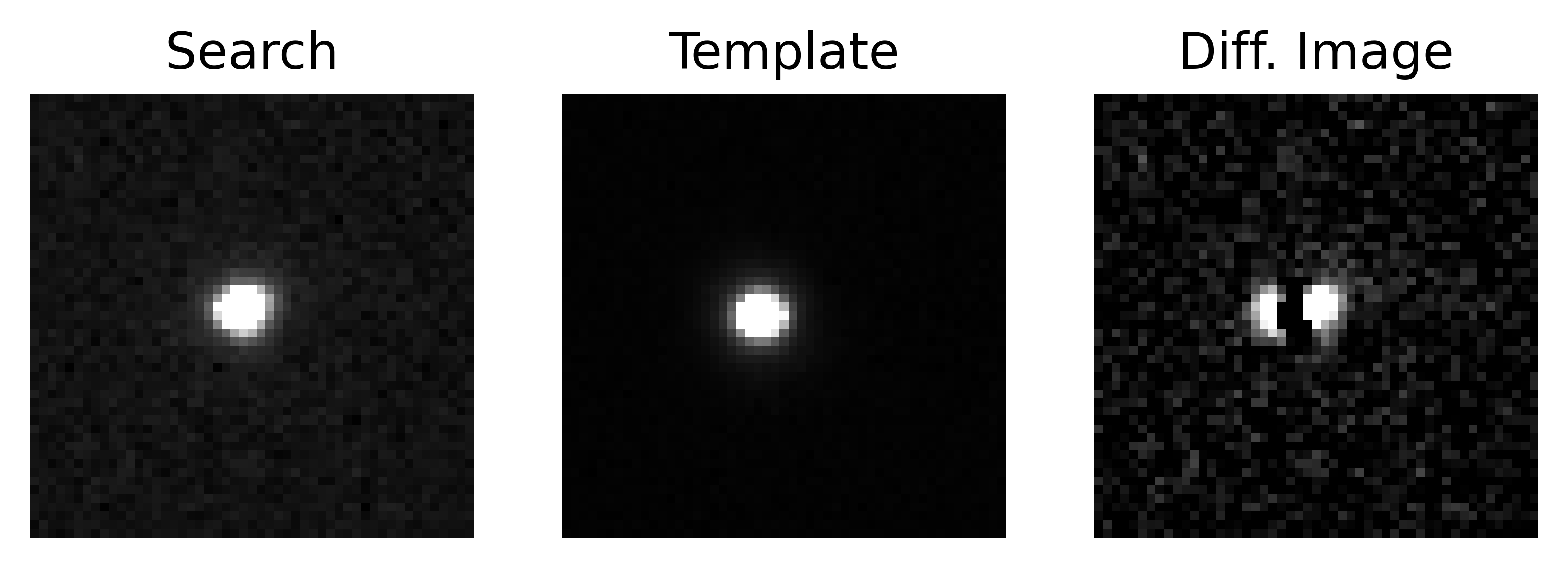}
    \includegraphics[width=\linewidth]{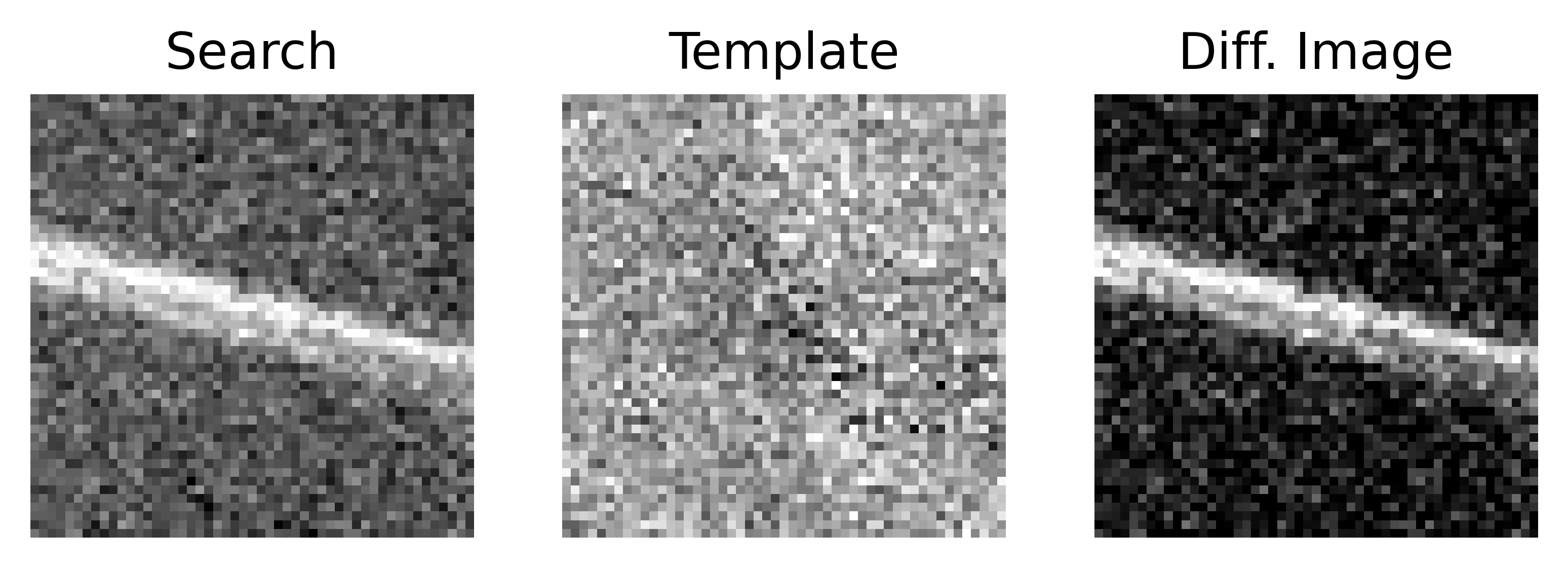}
    \centering
    \caption{{\it Top row}: A search image ({\it top left}), template image ({\it top middle}), and difference between the search and template image highlighting the position of a source of transient emission for a real transient source.  {\it Second row}: Same as the top, but for a transient embedded in a compact host galaxy.  {\it Third row}: A bad subtraction for a poorly aligned star, resulting in anomalous sources of emission in the difference image.  {\it Bottom row}: A diffraction spike that was not fully masked in the science image, resulting in an extended source of emission in the difference image.}
    \label{Fig:transient-examples}
\end{figure}

The dataset used in this section consists of a set of known real transients observed with T80 and processed by our pipeline, together with different kinds of artifacts produced by the same pipeline after difference imaging. We followed 20 different known supernovae (some of them in the same night) to have enough data to build a set of real transients. The cutout images produced by our transient detection pipeline are made using four SDSS-like $griz$ filters, having a cutout per filter. We collect three exposures per band, following the \splus\ Main Survey observing procedure \citep{MendesDeOliveira+19}, resulting in 12 different images for each transient candidate observation. Figure \ref{Fig:transient-examples} shows examples of different image cutouts at the end of our pipeline.

\begin{figure}
    \includegraphics[height=0.64\linewidth,width=\linewidth]{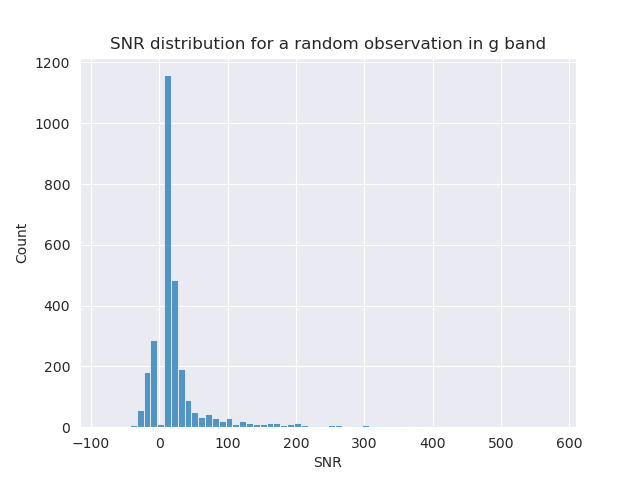}
    \includegraphics[height=0.64\linewidth,width=\linewidth]{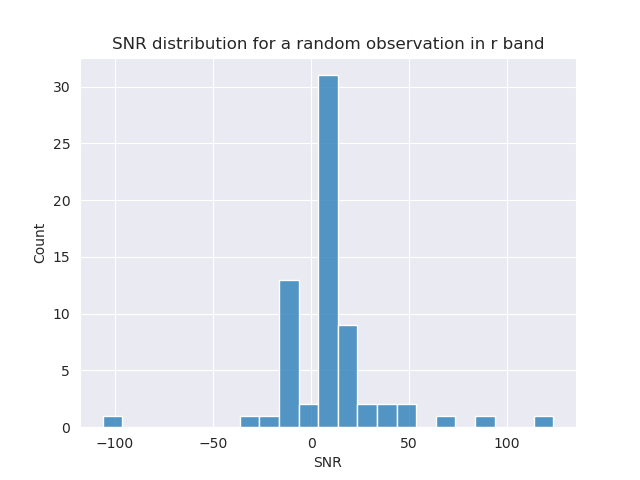}
    \includegraphics[height=0.64\linewidth,width=\linewidth]{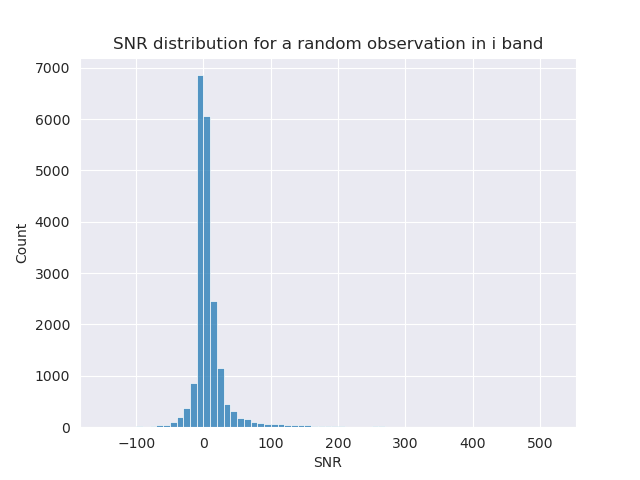}
    \includegraphics[height=0.64\linewidth,width=\linewidth]{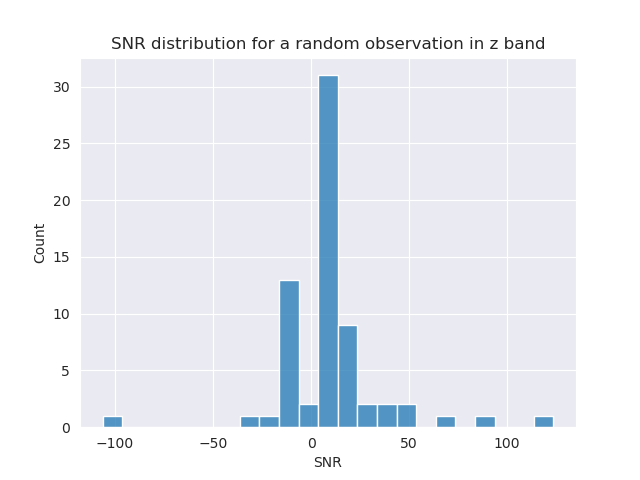}
    \centering
    \caption{SNR distribution per randomly selected image to build an artifact dataset for the machine learning algorithm.}
    \label{Fig:snr_dist}
\end{figure}

To build the artifact dataset, We randomly selected 4 images (one per filter) over the set of observations of all followed supernovae. From those, we made cutouts of all artifacts (excluding the real transient within the image), having a total sample of 10,000 samples. The Signal-to-noise (SNR) distribution of the total detection within the selected images is shown in figure \ref{Fig:snr_dist}. The supernovae dataset was built more straightforwardly. For each known followed transient, we made a cutout of all exposures per filter, having a total of 233 samples. We downsampled the artifact dataset to match the real transients and made use of data augmentation to have a final dataset.

Following the reduction procedure described in section 3, each image is processed with a search (or science, i.e., the original image from T80S), template (PS1, DECam, or SkyMapper image), and difference images (the subtraction between the two), all of which are simultaneously fed into the network through independent channels in the format of (51$\times$51$\times$3), similar to how the red, green, and blue image arrays that are fed into CNNs in traditional image analyses.  To ensure the integrity of our results, we avoid data leakage by assigning all 12 images of each observation to the training, validation, or test dataset. In this way, we prevent any bias in the model's test results that might occur if one of these images was used for training and another for testing. 

One of the challenges associated with the training and analysis of this dataset is that the transient candidates are contaminated by large numbers of bad subtractions. This resulted in an unbalanced classification problem. To address this issue, we downsampled the bad subtraction class to have an equal number of transient images. Finally, the total collection of images was partitioned approximately into three subsets, with 67\% of the data allocated for training, 13\%  for validation, and 20\% for testing. Finally, an additional set of 96 images was obtained to ensure our model's performance metrics.

\subsubsection{Pre-processing}

The deep learning methods are sensitive to visual features by construction, so we pre-process the data to enhance visualization, easing the training process. The pre-processing is divided into two main steps: contrast adjustment and normalization.

First, an image contrast adjustment was performed by clipping the image histogram, that is, we chose lower and upper limits and set every pixel above or below those values to be equal to the nearest bound. We choose the upper limited value to be the 99.2 \% percentile level for the search, template, and difference images. For the lower bound, we choose the 0.01 percentile. In other terms, every value above the upper limit (percentile 99.2\%) would be set to this threshold value, and every value below the lower bound/limit would be set to the percentile 1\% of the histogram values of each image. this is done for each channel independently to maintain the unique characteristics of each channel. 

Afterward, the images were normalized such that all pixels lie in the range [0,1], where for each image and each channel we subtract the minimum pixel value and divide by the maximum minus the minimum pixel value. This step improves the neural network training convergence. We made use of data augmentation techniques to enhance the training by applying rotations of 90, 180, and 270 degrees, horizontal flips, vertical flips, and combinations of rotations and flips. This increases the dataset by a factor of 8.

\subsubsection{Training}

A slight modification to the {\tt MobilNet} architecture was made by adding a flattening layer instead of the two last layers, which were originally an average pooling layer and a fully connected layer with one neuron \citep{Howard17}. The new architecture was designed to improve the generalization capability of the model.

The previously described model was trained using one of the
state-of-art optimizers: the {\tt Adam} \citep{Kingma14}, with a binary-cross entropy loss. We initialize the networks with pre-trained weights from the {\tt ImageNet} dataset to improve the computing time needed for convergence, overall performance, and stability of training, as observed in \citep{Bom21}.

The model was trained in a Multi-GPU cluster with 8 RTX 3090 with 24 GB of GPU memory each. The model training was implemented in {\tt TensorFlow2} \citep{TensorFlow}. To determine the optimal set of weights for the model, we train the network for an unlimited number of epochs until the Validation Loss value stops improving for 25 epochs. We selected the weights from the epoch with the lowest validation Loss, this method of weight selection maximizes performance on the validation set, which serves as a good indicator of how the model will perform on unseen data.

\subsection{Results}

In a binary classification problem, various metrics can be used to evaluate model performance, but for our particular problem of filtering artifacts while preserving transients, three main metrics are of particular interest: precision, recall, and false positive rate.

The precision (or purity) is defined as the proportion of true positive predictions made by the model among all positive predictions made. It represents the ability of the model to correctly identify true transients among all images classified as transients.  The recall (or completeness or true positive rate), is defined as the proportion of true positive predictions made by the model among all actual positives. it represents the ability of the model to correctly identify all true transients.
The false positive rate is the proportion of artifacts that are misclassified as transients.

The output of our CNN model is a score that can be interpreted as the probability of a given image being transient. Therefore, it is necessary to define a threshold, $t$, which is used to differentiate between true transients and artifacts. An image set is classified as a transient if the model's probability output exceeds $t$, and as an Artifact if the probability falls below $t$.  The threshold that maximizes the F1 score (the harmonic mean of precision and recall) is used as the operating threshold for determining positive and negative results.

\begin{figure}
    \includegraphics[width=\linewidth]{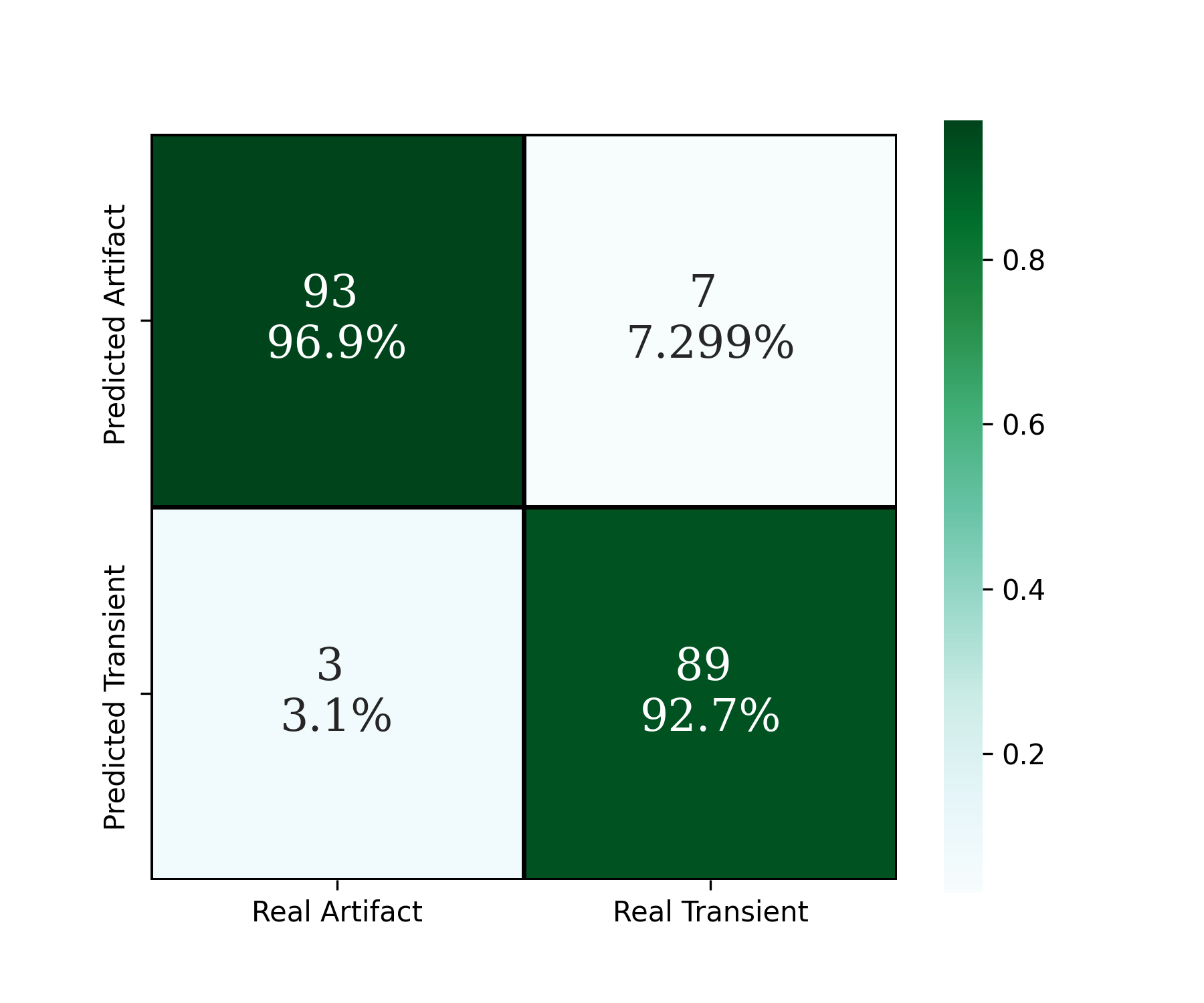}
    \centering
    \caption{Confusion Matrix for the CNN responsible for filtering possible artifacts on our final stages.}
    \label{fig:confusionmatrix}
\end{figure}

With this threshold, the false positive rate was determined to be 3.1\%, and the recall was 92.7\%.
This means that only 3.1\% of the images classified as transient were actual Artifacts, and 92.7\% of the true transient images were correctly identified as such. A confusion matrix was generated to further analyze the performance, which is shown in Figure~\ref{fig:confusionmatrix}.

\begin{figure}
    \includegraphics[width=\linewidth]{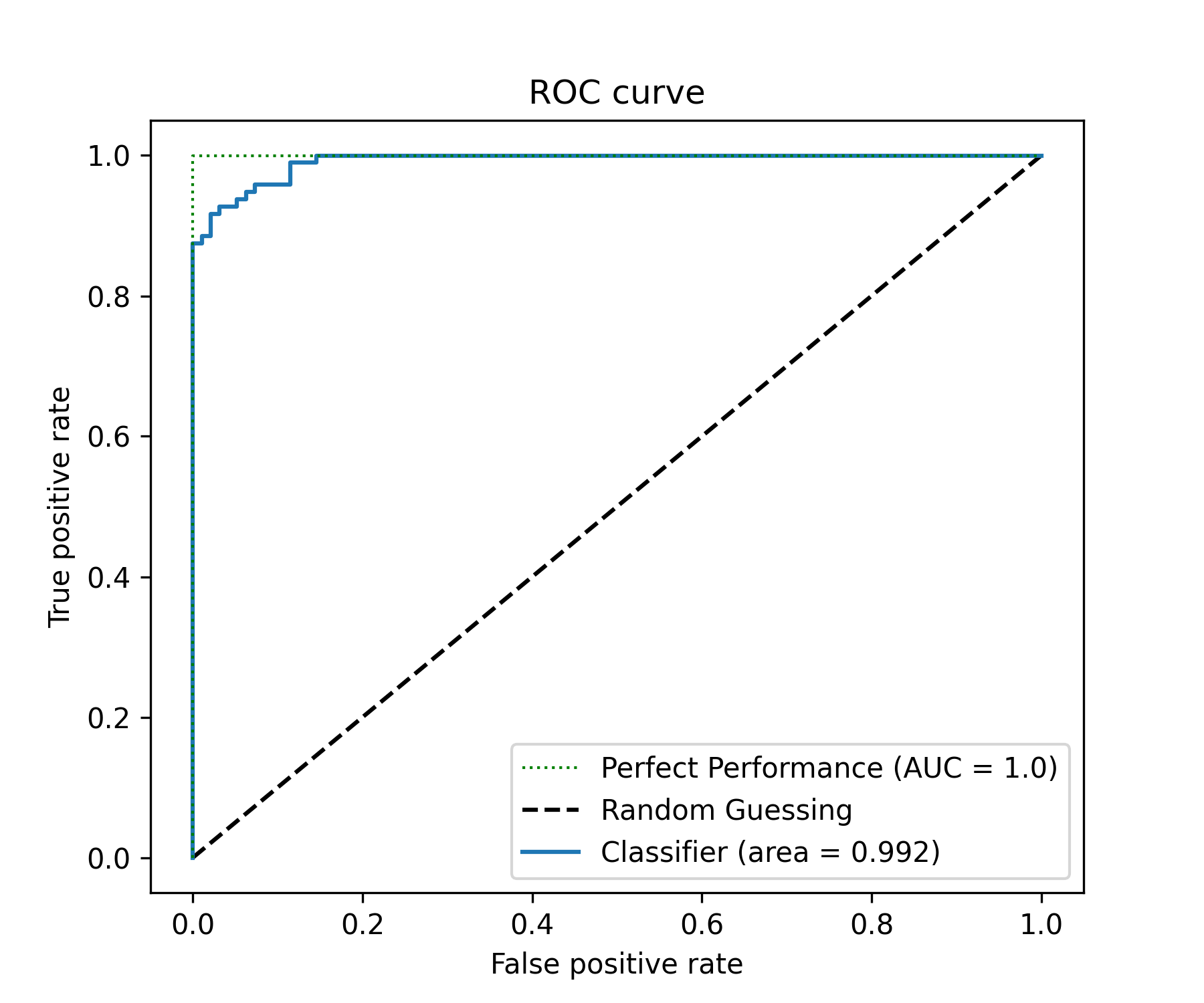}
    \caption{ Receiver operator characteristic curve for the trained network using the test sample. The area under the curve (AUC) is 0.992.}
    \label{fig:roc}
\end{figure}

\begin{figure}
    \includegraphics[width=\linewidth]{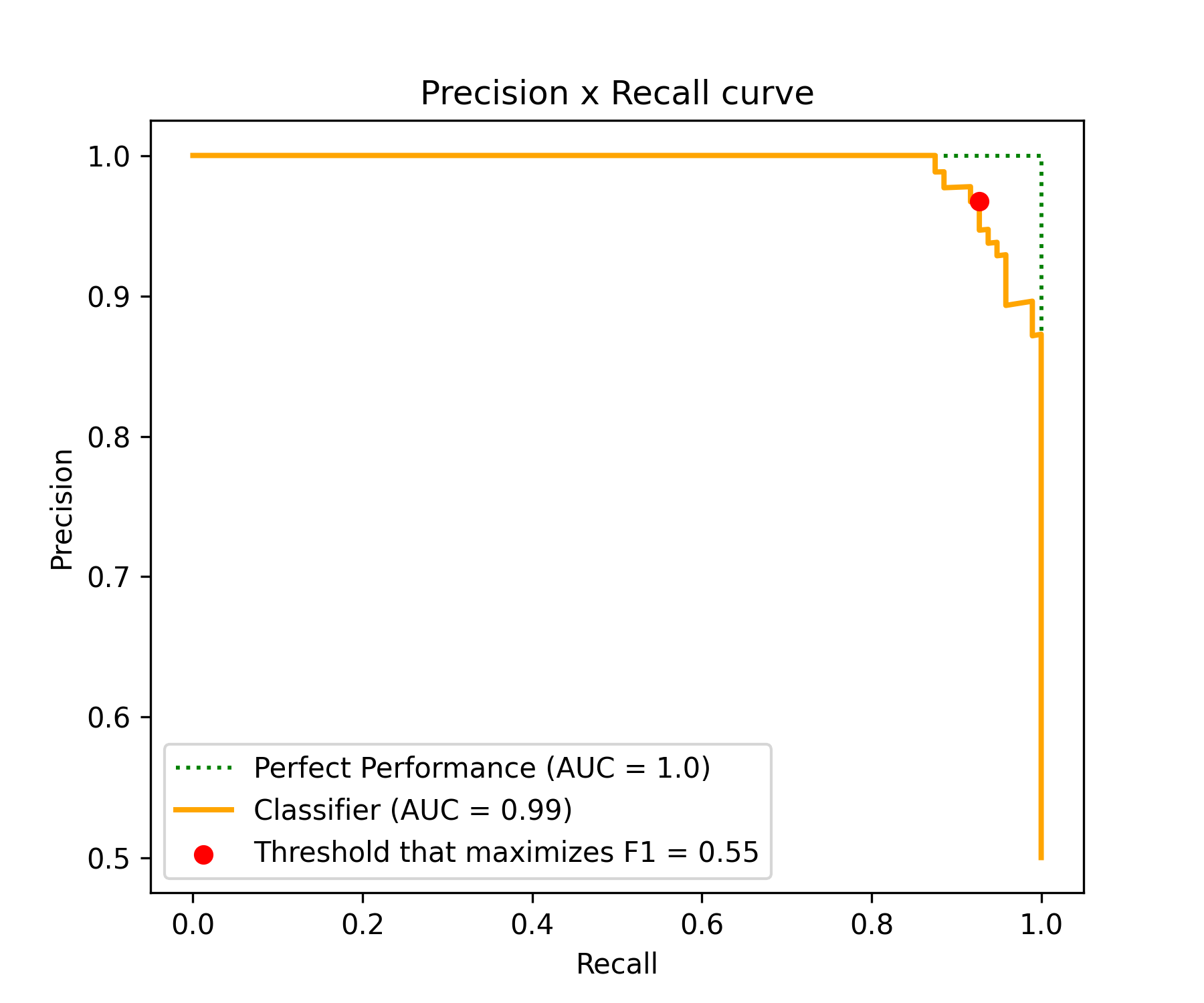}
    \caption{Precision-Recall (PR) Curve for the trained model using the test sample. The area under the curve (AUPRC) is 0.99.}
    \label{fig:precisionrecall}
\end{figure}

The Receiver Operating Characteristic (ROC)  curve shows the trade-off between the True Positive Rate (TPR) and False Positive Rate (FPR) of the model as the threshold is varied in the range [0, 1]. The Area
in the image refers to the area under the curve (AUC), our model achieved AUC$=$0.992 (Figure~\ref{fig:roc}).
Additionally, our model achieved an AUPRC (Area Under Precision-Recall Curve) of 0.99 (Figure~\ref{fig:precisionrecall}), which
aligns with the AUC from the ROC curve. This indicates the model’s capability to
accurately identify transients with high purity and maintain a high completeness. Such
results hold significant value in our analysis, particularly in scenarios where identifying
positive examples is crucial, such as detecting potential multimessengers.

\section{Extragalactic Transients Discovered in STEP Imaging}\label{sec:detected_transients}

After classifying  transient candidates using the flagging process described in Sections~\ref{sec:step_pipeline_overview} and \ref{sec:machine-learning}, we remove all likely image artifacts, asteroids, and variable stars before visually inspecting sources that we classify as likely extragalactic transients.  In particular, transients close to a source identified as a host galaxy \citep[i.e., with a photometric or spectroscopic redshift from \splus, Legacy, PS1, or the NASA/IPAC Extragalactic Database;][]{MendesDeOliveira+19,Zhou20,Beck21} or at high Galactic latitudes are inspected.

\begin{figure*}
    \centering
    \includegraphics[width=0.4\linewidth, height=0.5\linewidth]{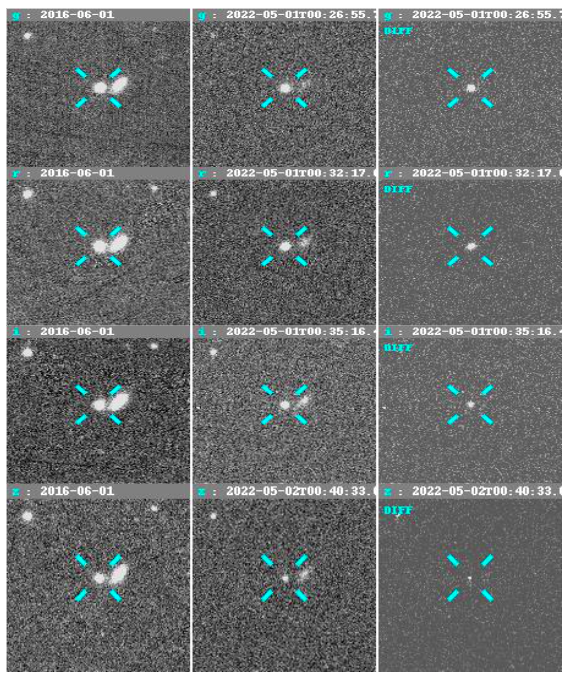}
    \caption{
\textit{From left to right}: Template image, Science Raw image, and difference image from our candidate at2022rri.
    }\label{fig:websniff}
\end{figure*}

Here we report the identification of two transients found through our analysis method described in Sections \ref{sec:step_pipeline_overview} and \ref{sec:machine-learning}. Our first supernova candidate detected within the S-PLUS Main Survey data was in 2022-05-01 at coordinates $\alpha$=10:31:09.221 and $\delta$=+00:19:36.25 (J2000) and reported to Transient Name Server (TNS) at 2022-08-21.\footnote{\url{https://www.wis-tns.org/object/2022rri}}. It was discovered at Magnitude $18.18 \pm 0.047$ (AB system, without MW Extinction correction) on $g$-Sloan band. Figure \ref{fig:websniff} shows our difference imaging data in the Sloan \textit{griz} bands for the candidate.

\begin{figure*}
    \centering
    \includegraphics[width=0.4\linewidth, height=0.5\linewidth]{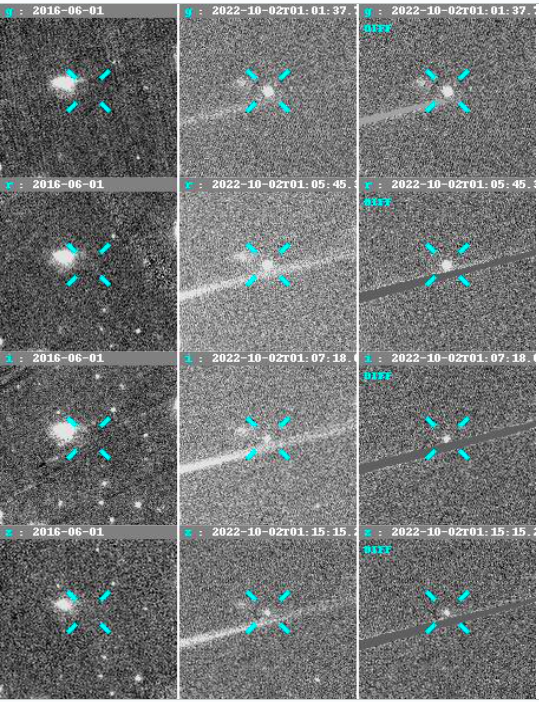}
    \caption{
	\label{fig:sn2022tiv}
\textit{From left to right}: Template image, science raw image, and difference image from our candidate. This image corresponds to SN 2022tiv.}
\end{figure*}

The second candidate found on our or list is the SN\,2022tiv\footnote{\url{https://www.wis-tns.org/object/2022tiv}}. It was initially reported by ATLAS \citep{2022tiv-disc} in 2022-09-02 at coordinates $\alpha$=21:23:57.450 and $\delta$=-22:44:07 (J2000) with magnitude 18.65 (AB System) in orange-ATLAS filter. It was rediscovered within our project 1 month after the S-PLUS Main Survey visited the same field. It was classified as a type Ia Supernova with redshift $z=0.03$ with spectrum data approximately 2 weeks after first reported \citep{2022TNSCR2660....1T}. Figure \ref{fig:sn2022tiv} shows our findings for this candidate.

\section{Supernova Follow-up}\label{sec:supernova}
During the observation run of this project we had followed supernovae that we flag as scientific interesting, either by its unique physical nature, rarirty or distance. We collect photometry data with 3 day cadence follow up of those targets with \splus\ using the same strategy as the Main Survey, when the time between the observations were available. We collected data for 4 different supernova: 2022crv\footnote{\url{https://www.wis-tns.org/object/2022crv}}, 2022ann\footnote{\url{https://www.wis-tns.org/object/2022ann}}, 2022qxu\footnote{\url{https://www.wis-tns.org/object/2022qxu}} and 2022acko\footnote{\url{https://www.wis-tns.org/object/2022acko}}. We summarize some information on those supernovae in Table \ref{tab:supernovae-targets} and follow-up photometry \citep[apart from SN\,2022ann, which was originally presented in][]{Davis22} from our pipeline is provided in appendix \ref{appendix:photometry}.

\begin{figure}
    \includegraphics[width=0.49\textwidth]{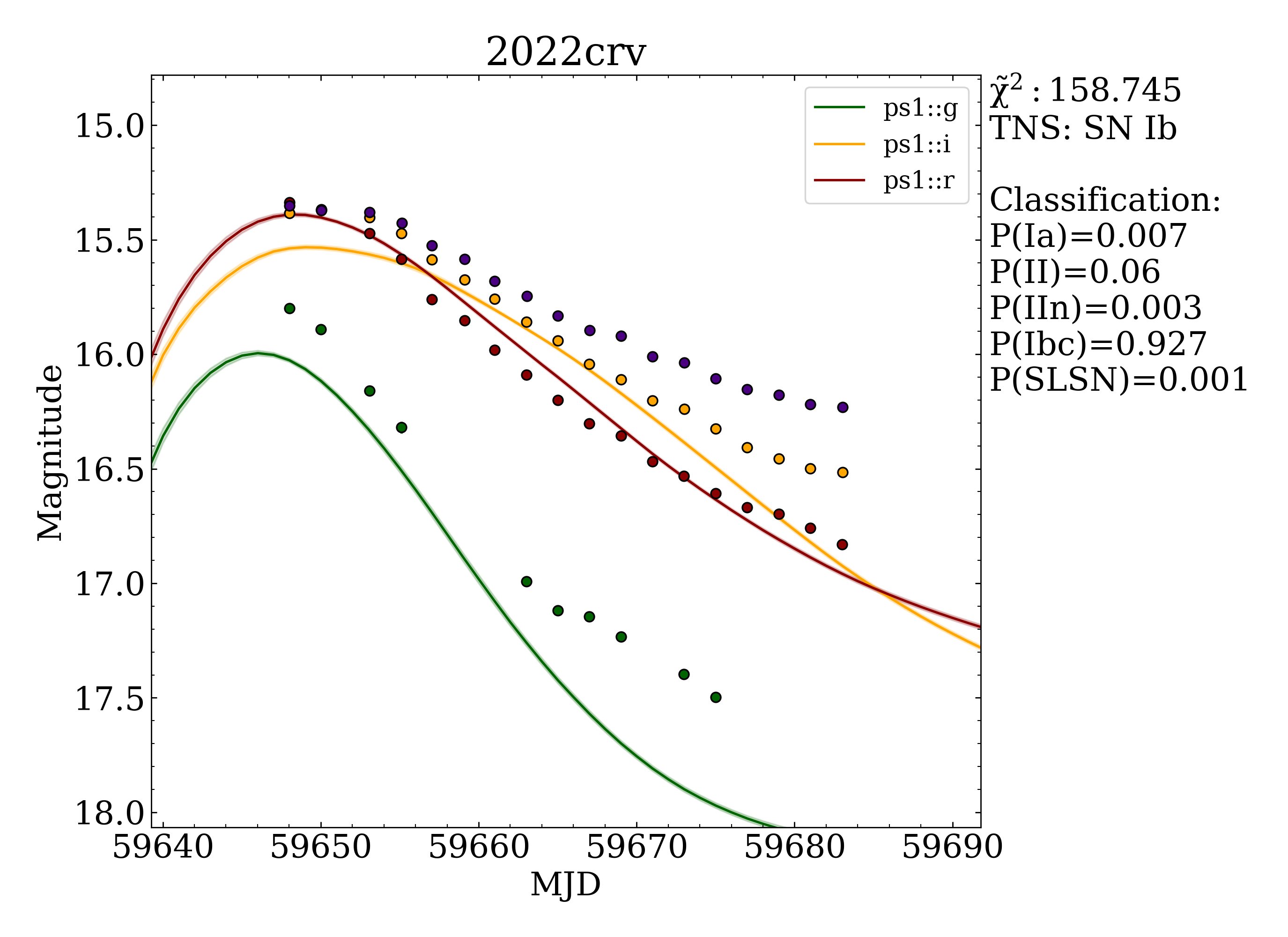}
    \includegraphics[width=0.49\textwidth]{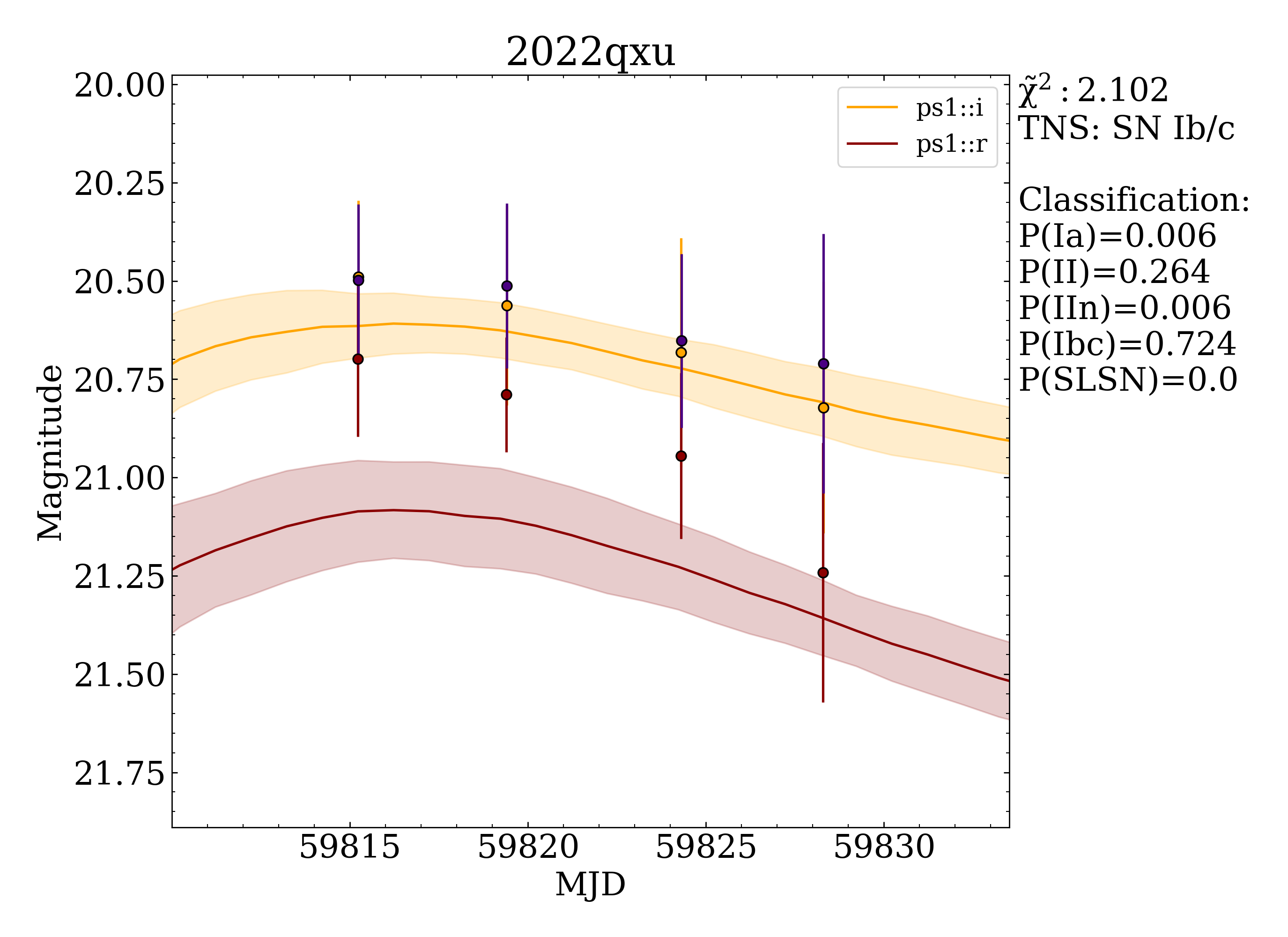}
    \includegraphics[width=0.49\textwidth]{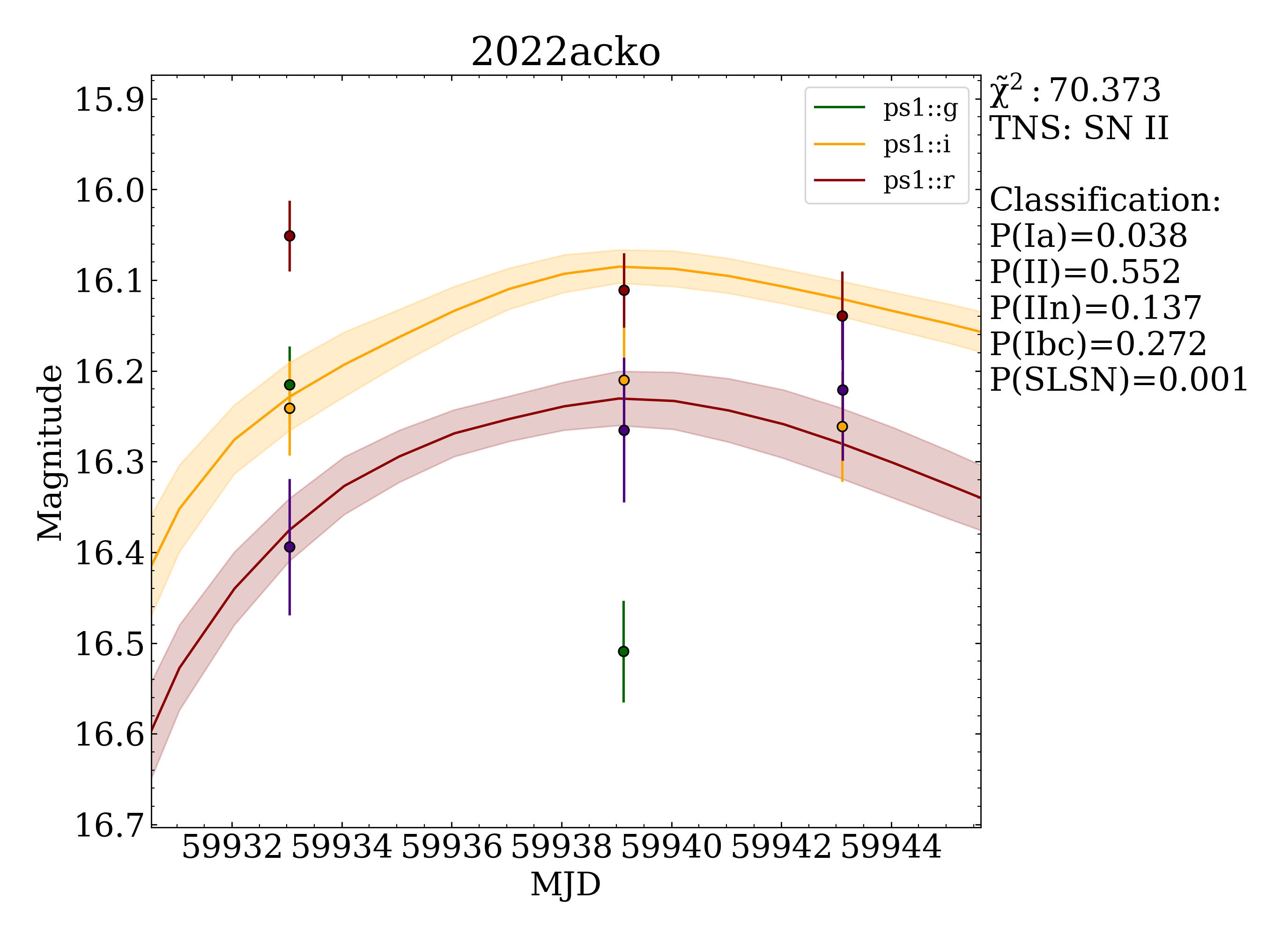}
    \caption{STEP light curves of SN\,2022crv, 2022qxu, and 2022acko as described in Section~\ref{sec:supernova}.  We photometrically classify our targets using methods implemented in \citep{Garretson23}.  In each case, the classifier accurately recovers the spectral type determined by optical spectroscopy and reports to the Transient Name Server.  All three sources were calibrated using the Pan-STARRS catalog, and so we use the classifications trained on Pan-STARRS photometry from \citep{Garretson23}, indicated by the ``ps1'' label for each band.}\label{fig:lc}
\end{figure}

We show the follow up light curves of SN\,2022crv, 2022qxu, and 2022acko in Fig.~\ref{fig:lc} with comparison to model light curves for photometric classification.  As a proof of concept, we utilize the  Recommender Engine For Intelligent Transient Tracking \citep[{\tt REFITT};][]{Sravan20} tool along with {\tt PanSNIP} \citep{Boone19} to photometrically classify our follow up light curves using methods implemented in \citep{Garretson23}.  Each of our targets has an easily identifiable host galaxy with a spectroscopic redshift, which we incorporate into our fit to aid in classification.  Although each target has a spectroscopic classification \citep[see][]{2022crv-class,2022qxu-class,2022acko-class}, we do not use any information from the transient other than follow-up data from STEP, including sky position and the in-band magnitudes reported in appendix \ref{appendix:photometry}.

The photometric classification scheme accurately recovers the SN type in each case despite the limited coverage of the overall light curve in some cases.  As described in \citep{Boone19} and \citep{Garretson23}, the combination of multi-band follow up, high cadences such as those from ZTF \citep{Bellm19} and STEP follow up, and host galaxy redshifts \citep[such as those from the S-PLUS Main Survey;][]{MendesDeOliveira+19} provide significant leverage in photometric classification, especially once classifiers are trained on full $ugrizy$ light curves.   In the era of Rubin, transient follow up surveys such as STEP can help increase the number of photometrically classified transients for cosmology, SN rates, and anomaly detection by providing high cadence ($<$5~day), multi-band photometry for the large number of transients discovered in the Legacy Survey of Space and Time. Since we can't cover the whole list of targets from LSST given our field-of-view capability, our future plans involve following selected targets, based on our science cases, and changing our untargeted survey to a targeted facility in the optical bands. Those targets can be selected by ranking tiles by the number of targets present on the fields, weighted with enhanced data we have at our disposal for further analysis.

\begin{table}
\centering
    \begin{tabular}{cccccc}
    \hline \hline
    Name & Host Galaxy & Redshift & $D_{L}$ \\
         &             &          & (Mpc) \\
    \hline \hline
     2022crv & NGC 3054 & 0.008091 & 40 \\
    2022ann & SDSS J101729.72-022535.6 & 0.04938 & 218 \\
    2022qxu & MCG -02-05-013 & 0.039964 &  177 \\
    2022acko & NGC1300 & 0.005264 & 23 \\
    \hline \hline
    \end{tabular}
    \caption{List of targeted supernovae during STEP DR1.}
    \label{tab:supernovae-targets}
\end{table}
\section{Strategy for Gravitational Wave Follow-up During LVK O4}\label{sec:gw_follow_up}

To perform a systematic search of optical counterparts of gravitational wave (GW) events, we design a strategy following the prescription presented by \citep{bom2023designing} using the same formalism. The strategy is defined as a set of parameters, namely the exposure times, broadband filter combinations g,r,i,z total probability area coverage, and time at the observation after the burst to detect the Kilonova with two independent passes over the GW probability region. For a given set of observational conditions parametrized by the effective exposure time factor, $t_{eff}$ a set of observational parameters $\Theta$ consisting of filters and exposure times, a choice of probability area coverage $\Omega$ in a moment $\tau$ after the GW alert, we calculate the probability of discovery, $P(\mathrm{discovery}|\hat{\Omega},\tau,t_\mathrm{eff},\Theta,\bar{z})\nonumber$ defined as:

\begin{align}
P_d &= \frac{\int_{\hat{\Omega}} d\Omega\,\,  d_{L}(\Omega)\, p(\Omega) }{\int_{\hat{\Omega}} d\Omega\,\, d_{L}(\Omega)}\, \times \sum_{j} p_{\alpha_j}\,,\label{eq:prob_area}
\end{align}

\noindent where $d_{L}$ is the luminosity distance and $p_{\alpha_j}$ are weights over the simulated light curves at redshift $z_i$ representing the best fit and its uncertainties \citep{Kilpatrick17} of the GW170817, described in terms of a Kasen kilonova model as a blue component \citep{Kasen17} \citep[see section 2.3 of ][for a description of the kN GW170817-like bright and blue and light curve model]{bom2023designing} that are within our limiting magnitude $m_{lim}(\Theta ,t_{eff})$. Similarly, the light curve simulations were performed by {\tt snana} \citep{snana}. We calculate the probability of confirmation $P_c$, i.e. the probability of two independent passes to detect the transient, this step is relevant to eliminate spurious detections during the GW counterpart search.  We define our strategy by allocating a maximum of $2$ nights per GW event.  In order to define a strategy, we use the same definitions from \citep{bom2023designing} to select a low telescope time strategy, i.e., a strategy that takes into account a balance between total telescope time used per night of observation while maintaining a high discovery probability. From a grid of possible exposure times ($60, 90, 120, 200, 300, 600, 1200, 3600$) and confidence level areas ($0.9, 0.8, 0.7$). We define and select a low telescope time strategy the one which uses the minimum telescope time around $10\%$ of the maximum probability of discovery from our simulations run. 

For well-localized events, one can describe the sky probability region of a gw event as an ellipse\footnote{See subsection Sky Localization Ellipse from \url{https://emfollow.docs.ligo.org/userguide/content.html}}. Each confidence interval is described as a smaller ellipse in comparison with an ellipse of higher confidence interval. In our strategy design, we left open the possibility of using deeper exposures on the core confidence levels 0.3 and 0.5 (or "inner" ellipses). We assume the observation can take a maximum of $8$ hours per night, $30s$ readout/slew time per pointing, and dark/grey night. It is also important to remark that our confirmation probability does not include day/night rhythm (we set a fixed time for night duration) and time lost for bad weather.

\subsection{Gravitational Wave Simulations}
For the GW simulations described in this section, we used the BAYESTAR\footnote{\url{https://lscsoft.docs.ligo.org/ligo.skymap/index.html}} \citep{PhysRevD.93.024013} ecosystem which uses LALSuite tools \citep{lalsuite}. The method used in this section to create the simulated events follows the work described in \citep{Petrov_2022}.

\begin{figure}
    \includegraphics[width=\linewidth]{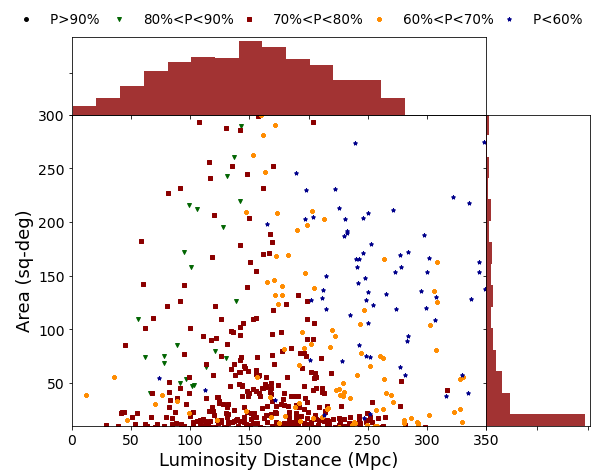}
    \centering
    \caption{The $90\%$ credible sky coverage distribution over luminosity distance for O4-like BNS detected events used in this paper. The different marks point to different Discovery probability ranges.}
    \label{Fig: gw_events}
\end{figure}

We begin our simulation with an injection of $10,000$ binary neutron stars (BNS) merger events with a TaylorF2 waveform limited to a maximum luminosity distance of 500 Mpc uniformly distributed over a sensitive volume where we presumed a Planck18 \citep{planckcolab} cosmology. The sensitivity functions for the Advanced LIGO, VIRGO, and KAGRA detectors were determined using the noise curve data available in the latest LIGO document \footnote{\url{https://dcc.ligo.org/LIGO-T2000012/public}}. All detectors have a duty cycle of $70\%$, following the discussions in \citep{LVK:prospects}. The BNS injections follow a Gaussian distribution for the masses, with mean $1.5$M$_\odot$ and standard deviation $1.1$M$_\odot$. We truncated the neutron star masses to be in the interval $1.1 < \mbox{M}_{\mbox{NS}} < 3.0$, taking in consideration the limits for Neutron Stars masses. The spins for both stars are uniformly distributed in the interval $[-0.05, 0.05]$ in the $z$ axis. After the injection, we use a Matched filter search to obtain detected events, using the same waveform in the injection process for reconstruction, resulting in 1076 detections. Here we define as detection those events which have a SNR $> 4$ in at least two detectors and a net SNR threshold > 12, where we added the measured SNR with a Gaussian noise. In the end, skymaps are produced using the \textbf{BAYESTAR} algorithm for the searched-out events. Figure \ref{Fig: gw_events} summarizes our findings, showing the $90\%$ credible area interval over the $90\%$ credible interval for Luminosity Distance, and the discovery probability of each simulated event.

\begin{figure}
   \includegraphics[width=\linewidth]{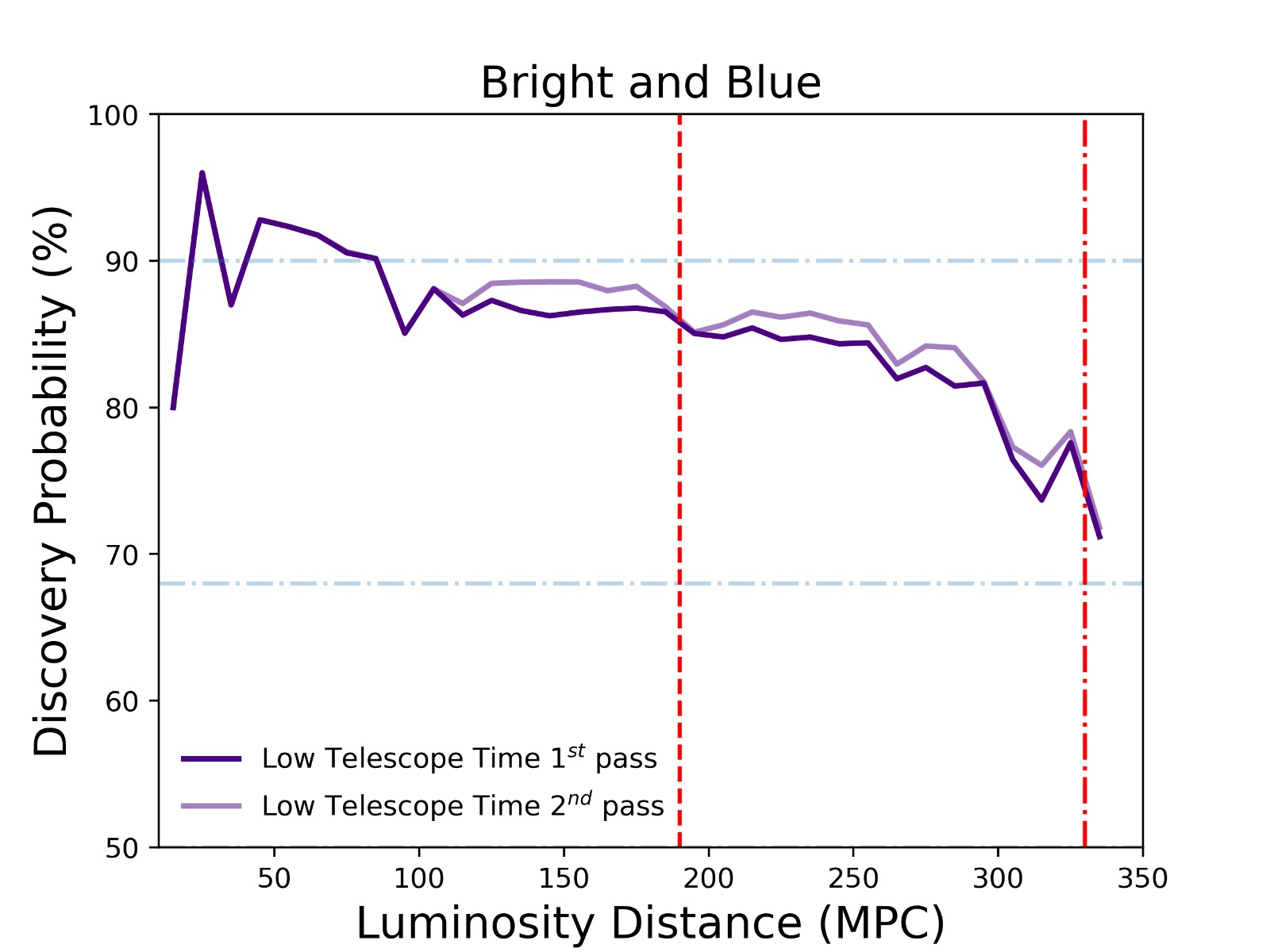}
   \centering
   \caption{Discovery Probability using a Low Telescope Time strategy with 2 passes. The first red vertical line at 190Mpc shows the current BNS Range sensitivity for LVK O4. The second line, at 330Mpc, is the limit for neutron star-black} hole range sensitivity.
   \label{fig:strategy}
\end{figure}

To properly design an observation strategy with the T80-South telescope, we selected events that have a sq-degree coverage in the $90\%$ credible region $< 300$ sq-deg. $791$ of the simulated events satisfies this condition. Given the Field of View from the telescope, we can cover  up to 100 sq-deg of the sky per night while saving telescope time using our method. The strategy adopted to search for an electromagnetic counterpart during LVK Fourth Observing run (hereafter, O4), follows closely the strategy designed in \citep{bom2023designing}, adapting the telescope specifics to our project. Figure \ref{fig:strategy} summarizes our discovery probability given a sample of BNS events, with simulated kilonovae light curves obtained with {\tt snana} \citep{snana}. The Low Telescope Time aims to increase the discovery probability while trying to minimize the amount of telescope time used in two passes.

It's important to remark that the simulations used in this work are based on assumptions about the planned sensitivity of LVK detectors. The predicted sensitives are close to those anticipated in mid-2022 for the O4 cycle. However, the current operation for O4 at the time of this writing diverges from the expected designed sensitivity. The Virgo detector, for example, hasn't joined the LIGO Handford and Livingston detectors from its beginning and remains uncertain when it will achieve the proposed sensitivity. The KAGRA detector will participate in O4 for a limited period, with a range for BNS merger below the estimated in this work (80Mpc). The estimate that comes from our analysis retains significant value, in terms of methodology and strategy for observation as an optimistic scenario in O4, although it does not reflect the reality of O4 so far.

 In the end, our strategy analysis outputs the best exposure time and filter that maximizes our chances of finding an electromagnetic counterpart. We further refine our tiles list using {\tt teglon} \citep{Kilpatrick21,Coulter23}, open-source software designed to optimally tile the LVK localization volume by crossmatching known galaxies in the GLADE galaxy catalog, under the assumption that the EM counterpart occurred in a galaxy within in that region.

\subsection{ Kilonovae Sensitivity}

We present our probability of discovery in our low-telescope time strategy in figure \ref{Fig: gw_events} which shows the area and distance of each event binned by its probability of discovery in the first pass. Considering our blue kN model with absolute magnitude in $i$ band peaks at $\sim -16$~\citep{bom2023designing},  most events have a discovery probability of $\sim 70-80$. Most of the strategies chosen by each event selected filters $r$ in both passes ($68\%$) or $g$ ($20\%$), or a combination of both $gr$ ($12\%$). The typical exposures selected by our optimization were $60s$ for the outer probability region and $90s$ in $37\%$($33\%$) of events in the first (second) pass. Another relevant configuration of exposure times was $200s$ or $300s$ for $22\%$($26\%$) of the core probability region in the first pass (second). The preferred configuration in terms of coverage was $90\%$ probability coverage and deeper exposures in the $30\%$ confidence level core.

\subsection{Candidate Vetting and Analysis}
To help mitigate the presence of false candidates, we employ a candidate vetting pipeline as outlined in section 3, which involves crossmatching our candidates with the minor planet center and transient name server, together with a neural network to help decrease false positives. Furthermore, conducting a secondary pass on the fields on a second night could aid in the identification and exclusion of other categories of transients. Whenever possible, we plan to validate candidates previously reported by the Zwicky Transient Facility (ZTF) via the FINK Broker API. Ultimately, we verify whether the targets in our final list have already been reported to the General Coordinates Network (GCN) and announce those that have not.

In the case of Binary Black Holes (BBHs) mergers, some models predict it's possible to find an electromagnetic counterpart as flares in Active Galactic Nuclei (AGNs) activities \citep[see, e.g.][]{rodríguezramírez2023optical,BartosRapid,tagawa23,graham23}. The flares from those events can last from days to weeks. We select a list of possible targets by crossmatching the 90\% credible region from BBHs mergers during O4 with AGN catalogs together with cuts on redshifts from possible host galaxies, decreasing the amount of sky area needed to be covered.

\section{Conclusions}\label{sec:conclusions}
\subsection{Summary}
The STEP Projects works as a transient survey in the Southern Hemisphere, using the capability of T80South Telescope with its 2 sq-deg field of view and the Main Survey observation strategy to report supernovae candidates to the astronomical community. Our project also works as a complementary source of photometric data of scientific interesting supernovae, helping understand the nature of those massive explosions. As in the era of multi-messenger astronomy,  STEP will follow up well-localized GW events that are inside T80 coverage, contributing to the discovery of kilonovae candidates and all the physics that comes within. Using a data reduction pipeline for difference imaging analysis, we successfully found and reported the transient AT\,2022rri and also recovered the known SN\,2022tiv when \splus\ was visiting the same field,  showing our capability in discovering new transients.  We estimate whether this is comparable to current wide-field surveys such as ZTF \citep{Bellm19} by noting that the average number of ZTF transient discoveries per area of sky surveyed is $\approx$0.016~transients~deg$^{-2}$.  Our survey covered 1272~deg$^{2}$ with a slightly shallower depth on average than ZTF of 0.5~mag, from which we estimate an expected $1272\times0.016\times2.512^{-0.5}=12.84\pm3.58$ transient discoveries.  Given that our discoveries are not at this rate yet, we expect that improvements in our template strategy, calibration, and machine learning methods will yield a much greater efficiency comparable to modern time-domain surveys.

\subsection{STEP as a GW Follow-up Program}\label{sec:gw_followup_summary}

With the current sensitivity on LVK detectors, new NS mergers are being discovered at distances far exceeding that of 40\,Mpc, consistent with the projection of a binary NS merger range $>$130~Mpc during O4 \citep{LVK:prospects}.  With the vastly increased search volume, many events will require coordinated follow up between multiple facilities to search the entire two-dimensional localization region to a depth consistent with kilonovae at the luminosity distance provided by the LVK.  Moreover, optimizing search strategy in real time requires candidate vetting and follow up and reporting on rapid timescales such that kilonovae can be rapidly identified and observed with other facilities.  Our GW strategy (Section~\ref{sec:gw_follow_up}) and participation in community-wide coordination efforts via NASA GCN \citep{2023GCN.33986} as well as the GW Treasure Map \citep{Wyatt20} will maximize the impact of STEP's search and follow up capabilities.

Throughout O4 and future LVK observing runs, the large localization regions of new compact object mergers will continue to tax the follow up capabilities of large footprint surveys such as ZTF \citep[which is also restricted to $\delta \gtrsim -30^{\circ}$;][]{Coughlin19,Graham20,Kasliwal20} and DECam \citep{Cowperthwaite17,soares2017electromagnetic,Andreoni19b,Goldstein19b,Garcia20,Morgan20}.  By coordinating T80S with these and other GW follow up facilities and optimizing for the science cases outlined above, we will maximize the impact of T80S observations for future multi-messenger events.

\subsection{The Next STEP in the Rubin Era}

With first light at the Rubin Observatory planned in 2024, thousands of new transients will be discovered per night reaching a typical per visit depth of $r=24.5$~mag \citep{LSST}.  However, the current nominal strategy will leave $\sim$20~day gaps in the single-filter light curves for Rubin-discovered transients and saturate on sources at $r\lesssim17$~mag.  This leaves significant room for small aperture, wide-field surveys to fill in the light curves of Rubin transients and observe those that peak at brighter magnitudes.

With a complementary filter set to Rubin and located at a similar observing site, STEP is ideally situated to obtain light curves of Rubin-discovered transients at $<$21~mag.  In particular, transients discovered within the S-PLUS main survey footprint have the benefit of multi-band host galaxy information and photometric redshifts to set the basic energy scale of the transient emission \citep{MendesDeOliveira+19}.  Combining this host galaxy information for photometric classification of transients has been shown to significantly improve the accuracy of transient classifiers \citep{Baldeschi20, Gagliano21}.  Although Rubin will eventually produce its own photometric redshift catalogs, this will take time and Rubin lacks narrow band UV filters used in S-PLUS that are optimized for photometric redshifts in the local Universe (i.e., at $z<0.1$).  By focusing on transient searches within the S-PLUS volume, STEP can leverage its ability to obtain high-cadence, multi-band light curves to maximize the science return for nearby Rubin-discovered transients.
\section*{Acknowledgments}

The S-PLUS project, including the T80-South robotic telescope and the S-PLUS scientific survey, was founded as a partnership between the Funda\c{c}\~{a}o de Amparo \`{a} Pesquisa do Estado de S\~{a}o Paulo (FAPESP), the Observat\'{o}rio Nacional (ON), the Federal University of Sergipe (UFS), and the Federal University of Santa Catarina (UFSC), with important financial and practical contributions from other collaborating institutes in Brazil, Chile (Universidad de La Serena), and Spain (Centro de Estudios de F\'{\i}sica del Cosmos de Arag\'{o}n, CEFCA). We further acknowledge financial support from the São Paulo Research Foundation (FAPESP), the Brazilian National Research Council (CNPq), the Coordination for the Improvement of Higher Education Personnel (CAPES), the Carlos Chagas Filho Rio de Janeiro State Research Foundation (FAPERJ), and the Brazilian Innovation Agency (FINEP).

Andre Santos and Clecio Bom acknowledges the financial support from CNPq (402577/2022-1)
Clecio Bom acknowledges the financial support from CNPq (316072/2021-4) and from FAPERJ (grants 201.456/2022 and 210.330/2022) and the FINEP contract 01.22.0505.00 (ref. 1891/22). The authors made use of Sci-Mind servers machines developed by the CBPF AI LAB team and would like to thank P. Russano and M. Portes de Albuquerque for all the support in infrastructure matters.

C.D. Kilpatrick acknowledges support from a CIERA postdoctoral fellowship.

F.R. Herpich acknowledges financial support from FAPESP (grant 2018/21661-9).

A. Alvarez-Candal acknowledges financial support from the Severo Ochoa grant CEX2021-001131-S funded by MCIN/AEI/ 10.13039/501100011033.

M.J. Sartori acknowledges financial support from FAPESP (grant 2022/00996-8).

E.A.D.L. acknowledges financial support from FAPESP (grant 2023/03688-5).

\section*{Data Availability}
The data from S-PLUS Main Survey will be available under data release in splus-cloud platform. The follow-up photometric data is released under the tables in appendix \ref{appendix:photometry}. The data related to gravitational wave simulations will be shared on reasonable request to the corresponding authors.
\bibliographystyle{mnras}
\bibliography{step} 

\begin{thebibliography}{}
\makeatletter
\relax
\def\mn@urlcharsother{\let\do\@makeother \do\$\do\&\do\#\do\^\do\_\do\%\do\~}
\def\mn@doi{\begingroup\mn@urlcharsother \@ifnextchar [ {\mn@doi@} {\mn@doi@[]}}
\def\mn@doi@[#1]#2{\def\@tempa{#1}\ifx\@tempa\@empty \href {http://dx.doi.org/#2} {doi:#2}\else \href {http://dx.doi.org/#2} {#1}\fi \endgroup}
\def\mn@eprint#1#2{\mn@eprint@#1:#2::\@nil}
\def\mn@eprint@arXiv#1{\href {http://arxiv.org/abs/#1} {{\tt arXiv:#1}}}
\def\mn@eprint@dblp#1{\href {http://dblp.uni-trier.de/rec/bibtex/#1.xml} {dblp:#1}}
\def\mn@eprint@#1:#2:#3:#4\@nil{\def\@tempa {#1}\def\@tempb {#2}\def\@tempc {#3}\ifx \@tempc \@empty \let \@tempc \@tempb \let \@tempb \@tempa \fi \ifx \@tempb \@empty \def\@tempb {arXiv}\fi \@ifundefined {mn@eprint@\@tempb}{\@tempb:\@tempc}{\expandafter \expandafter \csname mn@eprint@\@tempb\endcsname \expandafter{\@tempc}}}

\bibitem[\protect\citeauthoryear{Abadi et~al.,}{Abadi et~al.}{2015}]{TensorFlow}
Abadi M.,  et~al., 2015, {TensorFlow}: Large-Scale Machine Learning on Heterogeneous Systems, \url {https://www.tensorflow.org/}

\bibitem[\protect\citeauthoryear{{Abbott} et~al.,}{{Abbott} et~al.}{2017a}]{2017Natur.551...85A}
{Abbott} B.~P.,  et~al., 2017a, \mn@doi [\nat] {10.1038/nature24471}, \href {https://ui.adsabs.harvard.edu/abs/2017Natur.551...85A} {551, 85}

\bibitem[\protect\citeauthoryear{{Abbott} et~al.,}{{Abbott} et~al.}{2017b}]{Abbott17:mma}
{Abbott} B.~P.,  et~al., 2017b, \mn@doi [\apjl] {10.3847/2041-8213/aa91c9}, \href {https://ui.adsabs.harvard.edu/abs/2017ApJ...848L..12A} {848, L12}

\bibitem[\protect\citeauthoryear{{Abbott} et~al.,}{{Abbott} et~al.}{2020}]{LVK:prospects}
{Abbott} B.~P.,  et~al., 2020, \mn@doi [Living Reviews in Relativity] {10.1007/s41114-020-00026-9}, \href {https://ui.adsabs.harvard.edu/abs/2020LRR....23....3A} {23, 3}

\bibitem[\protect\citeauthoryear{{Ackley} et~al.,}{{Ackley} et~al.}{2020}]{Ackley20}
{Ackley} K.,  et~al., 2020, \mn@doi [\aap] {10.1051/0004-6361/202037669}, \href {https://ui.adsabs.harvard.edu/abs/2020A&A...643A.113A} {643, A113}

\bibitem[\protect\citeauthoryear{{Aleo} et~al.,}{{Aleo} et~al.}{2023}]{YSNE}
{Aleo} P.~D.,  et~al., 2023, \mn@doi [\apjs] {10.3847/1538-4365/acbfba}, \href {https://ui.adsabs.harvard.edu/abs/2023ApJS..266....9A} {266, 9}

\bibitem[\protect\citeauthoryear{{Alexander} et~al.,}{{Alexander} et~al.}{2021}]{Alexander21}
{Alexander} K.~D.,  et~al., 2021, arXiv e-prints, \href {https://ui.adsabs.harvard.edu/abs/2021arXiv210208957A} {p. arXiv:2102.08957}

\bibitem[\protect\citeauthoryear{{Almeida-Fernandes} et~al.,}{{Almeida-Fernandes} et~al.}{2022}]{AlmeidaFernandes+22}
{Almeida-Fernandes} F.,  et~al., 2022, \mn@doi [\mnras] {10.1093/mnras/stac284}, \href {https://ui.adsabs.harvard.edu/abs/2022MNRAS.511.4590A} {511, 4590}

\bibitem[\protect\citeauthoryear{{Andreoni} et~al.,}{{Andreoni} et~al.}{2019}]{Andreoni19b}
{Andreoni} I.,  et~al., 2019, \mn@doi [\apjl] {10.3847/2041-8213/ab3399}, \href {https://ui.adsabs.harvard.edu/abs/2019ApJ...881L..16A} {881, L16}

\bibitem[\protect\citeauthoryear{{Andreoni} et~al.,}{{Andreoni} et~al.}{2020}]{Andreoni19}
{Andreoni} I.,  et~al., 2020, \mn@doi [\apj] {10.3847/1538-4357/ab6a1b}, \href {https://ui.adsabs.harvard.edu/abs/2020ApJ...890..131A} {890, 131}

\bibitem[\protect\citeauthoryear{{Andreoni} et~al.,}{{Andreoni} et~al.}{2022a}]{Andreoni22}
{Andreoni} I.,  et~al., 2022a, \mn@doi [\apjs] {10.3847/1538-4365/ac617c}, \href {https://ui.adsabs.harvard.edu/abs/2022ApJS..260...18A} {260, 18}

\bibitem[\protect\citeauthoryear{Andreoni et~al.,}{Andreoni et~al.}{2022b}]{andreoni2022very}
Andreoni I.,  et~al., 2022b, Nature, 612, 430

\bibitem[\protect\citeauthoryear{{Andrews} et~al.,}{{Andrews} et~al.}{2022}]{2022crv-class}
{Andrews} J.~E.,  et~al., 2022, Transient Name Server Classification Report, \href {https://ui.adsabs.harvard.edu/abs/2022TNSCR.454....1A} {2022-454, 1}

\bibitem[\protect\citeauthoryear{{Antier} et~al.,}{{Antier} et~al.}{2020}]{Antier20}
{Antier} S.,  et~al., 2020, \mn@doi [\mnras] {10.1093/mnras/stz3142}, \href {https://ui.adsabs.harvard.edu/abs/2020MNRAS.492.3904A} {492, 3904}

\bibitem[\protect\citeauthoryear{{Arcavi}}{{Arcavi}}{2018}]{Arcavi18}
{Arcavi} I.,  2018, \mn@doi [\apjl] {10.3847/2041-8213/aab267}, \href {https://ui.adsabs.harvard.edu/abs/2018ApJ...855L..23A} {855, L23}

\bibitem[\protect\citeauthoryear{{Argast}, {Samland}, {Thielemann}  \& {Qian}}{{Argast} et~al.}{2004}]{Argast04}
{Argast} D.,  {Samland} M.,  {Thielemann} F.~K.,   {Qian} Y.~Z.,  2004, \mn@doi [\aap] {10.1051/0004-6361:20034265}, \href {https://ui.adsabs.harvard.edu/abs/2004A&A...416..997A} {416, 997}

\bibitem[\protect\citeauthoryear{{Arnould}, {Goriely}  \& {Takahashi}}{{Arnould} et~al.}{2007}]{Arnould07}
{Arnould} M.,  {Goriely} S.,   {Takahashi} K.,  2007, \mn@doi [\physrep] {10.1016/j.physrep.2007.06.002}, \href {https://ui.adsabs.harvard.edu/abs/2007PhR...450...97A} {450, 97}

\bibitem[\protect\citeauthoryear{{Baldeschi}, {Miller}, {Stroh}, {Margutti}  \& {Coppejans}}{{Baldeschi} et~al.}{2020}]{Baldeschi20}
{Baldeschi} A.,  {Miller} A.,  {Stroh} M.,  {Margutti} R.,   {Coppejans} D.~L.,  2020, \mn@doi [\apj] {10.3847/1538-4357/abb1c0}, \href {https://ui.adsabs.harvard.edu/abs/2020ApJ...902...60B} {902, 60}

\bibitem[\protect\citeauthoryear{{Bartos}, {Kocsis}, {Haiman}  \& {M{\'a}rka}}{{Bartos} et~al.}{2017}]{BartosRapid}
{Bartos} I.,  {Kocsis} B.,  {Haiman} Z.,   {M{\'a}rka} S.,  2017, \mn@doi [\apj] {10.3847/1538-4357/835/2/165}, \href {https://ui.adsabs.harvard.edu/abs/2017ApJ...835..165B} {835, 165}

\bibitem[\protect\citeauthoryear{{Beck}, {Szapudi}, {Flewelling}, {Holmberg}, {Magnier}  \& {Chambers}}{{Beck} et~al.}{2021}]{Beck21}
{Beck} R.,  {Szapudi} I.,  {Flewelling} H.,  {Holmberg} C.,  {Magnier} E.,   {Chambers} K.~C.,  2021, \mn@doi [\mnras] {10.1093/mnras/staa2587}, \href {https://ui.adsabs.harvard.edu/abs/2021MNRAS.500.1633B} {500, 1633}

\bibitem[\protect\citeauthoryear{{Becker}}{{Becker}}{2015}]{2015ascl.soft04004B}
{Becker} A.,  2015, {HOTPANTS: High Order Transform of PSF ANd Template Subtraction}, Astrophysics Source Code Library, record ascl:1504.004 (\mn@eprint {ascl} {1504.004})

\bibitem[\protect\citeauthoryear{{Belczynski}, {Kalogera}  \& {Bulik}}{{Belczynski} et~al.}{2002}]{Belczynski02}
{Belczynski} K.,  {Kalogera} V.,   {Bulik} T.,  2002, \mn@doi [\apj] {10.1086/340304}, \href {https://ui.adsabs.harvard.edu/abs/2002ApJ...572..407B} {572, 407}

\bibitem[\protect\citeauthoryear{{Bellm} et~al.,}{{Bellm} et~al.}{2019}]{Bellm19}
{Bellm} E.~C.,  et~al., 2019, \mn@doi [\pasp] {10.1088/1538-3873/aaecbe}, \href {https://ui.adsabs.harvard.edu/abs/2019PASP..131a8002B} {131, 018002}

\bibitem[\protect\citeauthoryear{{Bertin}}{{Bertin}}{2010}]{2010ascl.soft10068B}
{Bertin} E.,  2010, {SWarp: Resampling and Co-adding FITS Images Together}, Astrophysics Source Code Library, record ascl:1010.068 (\mn@eprint {ascl} {1010.068})

\bibitem[\protect\citeauthoryear{{Bertin} \& {Arnouts}}{{Bertin} \& {Arnouts}}{1996}]{1996A&AS..117..393B}
{Bertin} E.,  {Arnouts} S.,  1996, \mn@doi [\aaps] {10.1051/aas:1996164}, \href {https://ui.adsabs.harvard.edu/abs/1996A&AS..117..393B} {117, 393}

\bibitem[\protect\citeauthoryear{{Bom} \& {Palmese}}{{Bom} \& {Palmese}}{2023}]{bom23}
{Bom} C.~R.,  {Palmese} A.,  2023, \mn@doi [arXiv e-prints] {10.48550/arXiv.2307.01330}, \href {https://ui.adsabs.harvard.edu/abs/2023arXiv230701330B} {p. arXiv:2307.01330}

\bibitem[\protect\citeauthoryear{{Bom} et~al.,}{{Bom} et~al.}{2021}]{Bom21}
{Bom} C.~R.,  et~al., 2021, \mn@doi [\mnras] {10.1093/mnras/stab1981}, \href {https://ui.adsabs.harvard.edu/abs/2021MNRAS.507.1937B} {507, 1937}

\bibitem[\protect\citeauthoryear{Bom, Blanco~Valentin, Teles, Portes~de Albuquerque  \& Metcalf}{Bom et~al.}{2022}]{bom2022}
Bom C.,  Blanco~Valentin M.,  Teles K.,  Portes~de Albuquerque M.,   Metcalf R.~B.,  2022, Monthly Notices of the Royal Astronomical Society, 515, 5121

\bibitem[\protect\citeauthoryear{Bom et~al.,}{Bom et~al.}{2023}]{bom2023designing}
Bom C.~R.,  et~al., 2023, Designing an Optimal Kilonova Search using DECam for Gravitational Wave Events (\mn@eprint {arXiv} {2302.04878})

\bibitem[\protect\citeauthoryear{{Boone}}{{Boone}}{2019}]{Boone19}
{Boone} K.,  2019, \mn@doi [\aj] {10.3847/1538-3881/ab5182}, \href {https://ui.adsabs.harvard.edu/abs/2019AJ....158..257B} {158, 257}

\bibitem[\protect\citeauthoryear{Branch \& Tamman}{Branch \& Tamman}{1992}]{branch1992type}
Branch D.,  Tamman G.,  1992, Annual review of astronomy and astrophysics, 30, 359

\bibitem[\protect\citeauthoryear{Chen, Fishbach  \& Holz}{Chen et~al.}{2018}]{chen2018two}
Chen H.-Y.,  Fishbach M.,   Holz D.~E.,  2018, Nature, 562, 545

\bibitem[\protect\citeauthoryear{{Coughlin} et~al.,}{{Coughlin} et~al.}{2019}]{Coughlin19}
{Coughlin} M.~W.,  et~al., 2019, \mn@doi [\apjl] {10.3847/2041-8213/ab4ad8}, \href {https://ui.adsabs.harvard.edu/abs/2019ApJ...885L..19C} {885, L19}

\bibitem[\protect\citeauthoryear{{Coughlin} et~al.,}{{Coughlin} et~al.}{2020}]{Coughlin20b}
{Coughlin} M.~W.,  et~al., 2020, \mn@doi [\mnras] {10.1093/mnras/stz3457}, \href {https://ui.adsabs.harvard.edu/abs/2020MNRAS.492..863C} {492, 863}

\bibitem[\protect\citeauthoryear{{Coulter} et~al.,}{{Coulter} et~al.}{2017}]{Coulter17}
{Coulter} D.~A.,  et~al., 2017, \mn@doi [Science] {10.1126/science.aap9811}, \href {https://ui.adsabs.harvard.edu/abs/2017Sci...358.1556C} {358, 1556}

\bibitem[\protect\citeauthoryear{{Coulter}, {Kilpatrick}  \& {Foley}}{{Coulter} et~al.}{2023}]{Coulter23}
{Coulter} D.~A.,  {Kilpatrick} C.~D.,   {Foley} R.~J.,  2023, in preparation

\bibitem[\protect\citeauthoryear{{Cowperthwaite} et~al.,}{{Cowperthwaite} et~al.}{2017}]{Cowperthwaite17}
{Cowperthwaite} P.~S.,  et~al., 2017, \mn@doi [\apjl] {10.3847/2041-8213/aa8fc7}, \href {https://ui.adsabs.harvard.edu/abs/2017ApJ...848L..17C} {848, L17}

\bibitem[\protect\citeauthoryear{{Davis} et~al.,}{{Davis} et~al.}{2023}]{Davis22}
{Davis} K.~W.,  et~al., 2023, \mn@doi [\mnras] {10.1093/mnras/stad1433}, \href {https://ui.adsabs.harvard.edu/abs/2023MNRAS.523.2530D} {523, 2530}

\bibitem[\protect\citeauthoryear{{D{\'\i}az} et~al.,}{{D{\'\i}az} et~al.}{2017}]{Diaz17}
{D{\'\i}az} M.~C.,  et~al., 2017, \mn@doi [\apjl] {10.3847/2041-8213/aa9060}, \href {https://ui.adsabs.harvard.edu/abs/2017ApJ...848L..29D} {848, L29}

\bibitem[\protect\citeauthoryear{{Dobie} et~al.,}{{Dobie} et~al.}{2019}]{Dobie19}
{Dobie} D.,  et~al., 2019, \mn@doi [\apjl] {10.3847/2041-8213/ab59db}, \href {https://ui.adsabs.harvard.edu/abs/2019ApJ...887L..13D} {887, L13}

\bibitem[\protect\citeauthoryear{{Dominik} et~al.,}{{Dominik} et~al.}{2015}]{Dominik15}
{Dominik} M.,  et~al., 2015, \mn@doi [\apj] {10.1088/0004-637X/806/2/263}, \href {https://ui.adsabs.harvard.edu/abs/2015ApJ...806..263D} {806, 263}

\bibitem[\protect\citeauthoryear{{Duev} et~al.,}{{Duev} et~al.}{2019}]{Duev19}
{Duev} D.~A.,  et~al., 2019, \mn@doi [\mnras] {10.1093/mnras/stz2357}, \href {https://ui.adsabs.harvard.edu/abs/2019MNRAS.489.3582D} {489, 3582}

\bibitem[\protect\citeauthoryear{{Eichler}, {Livio}, {Piran}  \& {Schramm}}{{Eichler} et~al.}{1989}]{Eichler89}
{Eichler} D.,  {Livio} M.,  {Piran} T.,   {Schramm} D.~N.,  1989, \mn@doi [\nat] {10.1038/340126a0}, \href {https://ui.adsabs.harvard.edu/abs/1989Natur.340..126E} {340, 126}

\bibitem[\protect\citeauthoryear{{Filippenko}}{{Filippenko}}{1992}]{Filippenko92}
{Filippenko} A.~V.,  1992, in {Filippenko} A.~V.,  ed.,  Astronomical Society of the Pacific Conference Series Vol. 103, Robotic Telescopes in the 1990s. pp 55--66

\bibitem[\protect\citeauthoryear{{Filippenko}}{{Filippenko}}{2005}]{Filippenko05}
{Filippenko} A.~V.,  2005, in {Humphreys} R.,  {Stanek} K.,  eds,  Astronomical Society of the Pacific Conference Series Vol. 332, The Fate of the Most Massive Stars. p.~34 (\mn@eprint {arXiv} {astro-ph/0412029}), \mn@doi{10.48550/arXiv.astro-ph/0412029}

\bibitem[\protect\citeauthoryear{{Finkbeiner} et~al.,}{{Finkbeiner} et~al.}{2016}]{Finkbeiner16}
{Finkbeiner} D.~P.,  et~al., 2016, \mn@doi [\apj] {10.3847/0004-637X/822/2/66}, \href {https://ui.adsabs.harvard.edu/abs/2016ApJ...822...66F} {822, 66}

\bibitem[\protect\citeauthoryear{{F{\"o}rster} et~al.,}{{F{\"o}rster} et~al.}{2021}]{Forster21}
{F{\"o}rster} F.,  et~al., 2021, \mn@doi [\aj] {10.3847/1538-3881/abe9bc}, \href {https://ui.adsabs.harvard.edu/abs/2021AJ....161..242F} {161, 242}

\bibitem[\protect\citeauthoryear{{Gagliano}, {Narayan}, {Engel}, {Carrasco Kind}  \& {LSST Dark Energy Science Collaboration}}{{Gagliano} et~al.}{2021}]{Gagliano21}
{Gagliano} A.,  {Narayan} G.,  {Engel} A.,  {Carrasco Kind} M.,   {LSST Dark Energy Science Collaboration} 2021, \mn@doi [\apj] {10.3847/1538-4357/abd02b}, \href {https://ui.adsabs.harvard.edu/abs/2021ApJ...908..170G} {908, 170}

\bibitem[\protect\citeauthoryear{{Gaia Collaboration} et~al.,}{{Gaia Collaboration} et~al.}{2023}]{GaiaDR3}
{Gaia Collaboration} et~al., 2023, \mn@doi [\aap] {10.1051/0004-6361/202243940}, \href {https://ui.adsabs.harvard.edu/abs/2023A&A...674A...1G} {674, A1}

\bibitem[\protect\citeauthoryear{{Gal-Yam} et~al.,}{{Gal-Yam} et~al.}{2009}]{Gal-Yam09}
{Gal-Yam} A.,  et~al., 2009, \mn@doi [\nat] {10.1038/nature08579}, \href {https://ui.adsabs.harvard.edu/abs/2009Natur.462..624G} {462, 624}

\bibitem[\protect\citeauthoryear{{Garcia} et~al.,}{{Garcia} et~al.}{2020}]{Garcia20}
{Garcia} A.,  et~al., 2020, \mn@doi [\apj] {10.3847/1538-4357/abb823}, \href {https://ui.adsabs.harvard.edu/abs/2020ApJ...903...75G} {903, 75}

\bibitem[\protect\citeauthoryear{{Garretson} \& {Milisavljevic}}{{Garretson} \& {Milisavljevic}}{2023}]{Garretson23}
{Garretson} B.,  {Milisavljevic} D.,  2023, in prep.

\bibitem[\protect\citeauthoryear{{Goldstein} et~al.,}{{Goldstein} et~al.}{2019}]{Goldstein19b}
{Goldstein} D.~A.,  et~al., 2019, \mn@doi [\apjl] {10.3847/2041-8213/ab3046}, \href {https://ui.adsabs.harvard.edu/abs/2019ApJ...881L...7G} {881, L7}

\bibitem[\protect\citeauthoryear{{Gomez} et~al.,}{{Gomez} et~al.}{2019}]{Gomez19}
{Gomez} S.,  et~al., 2019, \mn@doi [\apjl] {10.3847/2041-8213/ab4ad5}, \href {https://ui.adsabs.harvard.edu/abs/2019ApJ...884L..55G} {884, L55}

\bibitem[\protect\citeauthoryear{{Graham} et~al.,}{{Graham} et~al.}{2020}]{Graham20}
{Graham} M.~J.,  et~al., 2020, \mn@doi [\prl] {10.1103/PhysRevLett.124.251102}, \href {https://ui.adsabs.harvard.edu/abs/2020PhRvL.124y1102G} {124, 251102}

\bibitem[\protect\citeauthoryear{{Graham} et~al.,}{{Graham} et~al.}{2023}]{graham23}
{Graham} M.~J.,  et~al., 2023, \mn@doi [\apj] {10.3847/1538-4357/aca480}, \href {https://ui.adsabs.harvard.edu/abs/2023ApJ...942...99G} {942, 99}

\bibitem[\protect\citeauthoryear{{Hamuy} et~al.,}{{Hamuy} et~al.}{1993}]{Hamuy93}
{Hamuy} M.,  et~al., 1993, \mn@doi [\aj] {10.1086/116811}, \href {https://ui.adsabs.harvard.edu/abs/1993AJ....106.2392H} {106, 2392}

\bibitem[\protect\citeauthoryear{{He}, {Zhang}, {Ren}  \& {Sun}}{{He} et~al.}{2015}]{He2016}
{He} K.,  {Zhang} X.,  {Ren} S.,   {Sun} J.,  2015, 2016 IEEE Conference on Computer Vision and Pattern Recognition

\bibitem[\protect\citeauthoryear{{Hosseinzadeh} et~al.,}{{Hosseinzadeh} et~al.}{2019}]{Hosseinzadeh19}
{Hosseinzadeh} G.,  et~al., 2019, \mn@doi [\apjl] {10.3847/2041-8213/ab271c}, \href {https://ui.adsabs.harvard.edu/abs/2019ApJ...880L...4H} {880, L4}

\bibitem[\protect\citeauthoryear{{Howard}}{{Howard}}{2017}]{Howard17}
{Howard} E.~M.,  2017, in {Lorente} N.~P.~F.,  {Shortridge} K.,   {Wayth} R.,  eds,  Astronomical Society of the Pacific Conference Series Vol. 512, Astronomical Data Analysis Software and Systems XXV. p.~245

\bibitem[\protect\citeauthoryear{{Hubble}}{{Hubble}}{1929}]{1929PNAS...15..168H}
{Hubble} E.,  1929, \mn@doi [Proceedings of the National Academy of Science] {10.1073/pnas.15.3.168}, \href {https://ui.adsabs.harvard.edu/abs/1929PNAS...15..168H} {15, 168}

\bibitem[\protect\citeauthoryear{{Ivezi{\'c}} et~al.,}{{Ivezi{\'c}} et~al.}{2019}]{LSST}
{Ivezi{\'c}} {\v{Z}}.,  et~al., 2019, \mn@doi [\apj] {10.3847/1538-4357/ab042c}, \href {https://ui.adsabs.harvard.edu/abs/2019ApJ...873..111I} {873, 111}

\bibitem[\protect\citeauthoryear{{Izzo}, {Langeroodi}  \& {Davis}}{{Izzo} et~al.}{2022}]{2022qxu-class}
{Izzo} L.,  {Langeroodi} D.,   {Davis} K.,  2022, Transient Name Server Classification Report, \href {https://ui.adsabs.harvard.edu/abs/2022TNSCR2337....1I} {2022-2337, 1}

\bibitem[\protect\citeauthoryear{{Kaiser} et~al.,}{{Kaiser} et~al.}{2002}]{2002SPIE.4836..154K}
{Kaiser} N.,  et~al., 2002, in {Tyson} J.~A.,  {Wolff} S.,  eds,  Society of Photo-Optical Instrumentation Engineers (SPIE) Conference Series Vol. 4836, Survey and Other Telescope Technologies and Discoveries. pp 154--164, \mn@doi{10.1117/12.457365}

\bibitem[\protect\citeauthoryear{{Karambelkar} et~al.,}{{Karambelkar} et~al.}{2023}]{Karambelkar23}
{Karambelkar} V.~R.,  et~al., 2023, \mn@doi [\apj] {10.3847/1538-4357/acc2b9}, \href {https://ui.adsabs.harvard.edu/abs/2023ApJ...948..137K} {948, 137}

\bibitem[\protect\citeauthoryear{{Kasen}, {Metzger}, {Barnes}, {Quataert}  \& {Ramirez-Ruiz}}{{Kasen} et~al.}{2017}]{Kasen17}
{Kasen} D.,  {Metzger} B.,  {Barnes} J.,  {Quataert} E.,   {Ramirez-Ruiz} E.,  2017, \mn@doi [\nat] {10.1038/nature24453}, \href {https://ui.adsabs.harvard.edu/abs/2017Natur.551...80K} {551, 80}

\bibitem[\protect\citeauthoryear{{Kasliwal} et~al.,}{{Kasliwal} et~al.}{2020}]{Kasliwal20}
{Kasliwal} M.~M.,  et~al., 2020, \mn@doi [\apj] {10.3847/1538-4357/abc335}, \href {https://ui.adsabs.harvard.edu/abs/2020ApJ...905..145K} {905, 145}

\bibitem[\protect\citeauthoryear{{Kessler} et~al.,}{{Kessler} et~al.}{2009}]{snana}
{Kessler} R.,  et~al., 2009, \mn@doi [\pasp] {10.1086/605984}, \href {https://ui.adsabs.harvard.edu/abs/2009PASP..121.1028K} {121, 1028}

\bibitem[\protect\citeauthoryear{Kilpatrick}{Kilpatrick}{2023}]{candidates}
Kilpatrick C.~D.,  2023, charliekilpatrick/candidates: candidates v1.0, \mn@doi{10.5281/zenodo.8172608}, \url {https://doi.org/10.5281/zenodo.8172608}

\bibitem[\protect\citeauthoryear{{Kilpatrick} et~al.,}{{Kilpatrick} et~al.}{2017}]{Kilpatrick17}
{Kilpatrick} C.~D.,  et~al., 2017, \mn@doi [Science] {10.1126/science.aaq0073}, \href {https://ui.adsabs.harvard.edu/abs/2017Sci...358.1583K} {358, 1583}

\bibitem[\protect\citeauthoryear{{Kilpatrick} et~al.,}{{Kilpatrick} et~al.}{2021}]{Kilpatrick21}
{Kilpatrick} C.~D.,  et~al., 2021, \mn@doi [\apj] {10.3847/1538-4357/ac23c6}, \href {https://ui.adsabs.harvard.edu/abs/2021ApJ...923..258K} {923, 258}

\bibitem[\protect\citeauthoryear{Kingma \& Ba}{Kingma \& Ba}{2014}]{Kingma14}
Kingma D.~P.,  Ba J.,  2014, arXiv preprint arXiv:1412.6980

\bibitem[\protect\citeauthoryear{{Kulkarni} et~al.,}{{Kulkarni} et~al.}{2007}]{Kulkarni07}
{Kulkarni} S.~R.,  et~al., 2007, \mn@doi [\nat] {10.1038/nature05822}, \href {https://ui.adsabs.harvard.edu/abs/2007Natur.447..458K} {447, 458}

\bibitem[\protect\citeauthoryear{{LIGO Scientific Collaboration}}{{LIGO Scientific Collaboration}}{2018}]{lalsuite}
{LIGO Scientific Collaboration} 2018, {LIGO} {A}lgorithm {L}ibrary - {LALS}uite, free software (GPL), \mn@doi{10.7935/GT1W-FZ16}

\bibitem[\protect\citeauthoryear{{LSST Science Collaboration} et~al.,}{{LSST Science Collaboration} et~al.}{2009}]{LSST09}
{LSST Science Collaboration} et~al., 2009, \mn@doi [arXiv e-prints] {10.48550/arXiv.0912.0201}, \href {https://ui.adsabs.harvard.edu/abs/2009arXiv0912.0201L} {p. arXiv:0912.0201}

\bibitem[\protect\citeauthoryear{{Lang}, {Hogg}, {Mierle}, {Blanton}  \& {Roweis}}{{Lang} et~al.}{2010}]{astrometry.net}
{Lang} D.,  {Hogg} D.~W.,  {Mierle} K.,  {Blanton} M.,   {Roweis} S.,  2010, \mn@doi [\aj] {10.1088/0004-6256/139/5/1782}, \href {https://ui.adsabs.harvard.edu/abs/2010AJ....139.1782L} {139, 1782}

\bibitem[\protect\citeauthoryear{{Li}, {Cai}, {Zhai}, {Zhang}  \& {Wang}}{{Li} et~al.}{2022}]{2022acko-class}
{Li} L.,  {Cai} Y.,  {Zhai} Q.,  {Zhang} J.,   {Wang} X.,  2022, Transient Name Server Classification Report, \href {https://ui.adsabs.harvard.edu/abs/2022TNSCR3549....1L} {2022-3549, 1}

\bibitem[\protect\citeauthoryear{{Lundquist} et~al.,}{{Lundquist} et~al.}{2019}]{Lundquist19}
{Lundquist} M.~J.,  et~al., 2019, \mn@doi [\apjl] {10.3847/2041-8213/ab32f2}, \href {https://ui.adsabs.harvard.edu/abs/2019ApJ...881L..26L} {881, L26}

\bibitem[\protect\citeauthoryear{{MacFadyen}, {Woosley}  \& {Heger}}{{MacFadyen} et~al.}{2001}]{MacFadyen01}
{MacFadyen} A.~I.,  {Woosley} S.~E.,   {Heger} A.,  2001, \mn@doi [\apj] {10.1086/319698}, \href {https://ui.adsabs.harvard.edu/abs/2001ApJ...550..410M} {550, 410}

\bibitem[\protect\citeauthoryear{{Maza}}{{Maza}}{1980}]{1980tsup.work....7M}
{Maza} J.,  1980, in {Wheeler} J.~C.,  ed., Texas Workshop on Type I Supernovae. p.~7

\bibitem[\protect\citeauthoryear{{McQuillan}, {Aigrain}  \& {Roberts}}{{McQuillan} et~al.}{2012}]{McQuillan12}
{McQuillan} A.,  {Aigrain} S.,   {Roberts} S.,  2012, \mn@doi [\aap] {10.1051/0004-6361/201016148}, \href {https://ui.adsabs.harvard.edu/abs/2012A&A...539A.137M} {539, A137}

\bibitem[\protect\citeauthoryear{{Mendes de Oliveira} et~al.,}{{Mendes de Oliveira} et~al.}{2019}]{MendesDeOliveira+19}
{Mendes de Oliveira} C.,  et~al., 2019, \mn@doi [\mnras] {10.1093/mnras/stz1985}, \href {https://ui.adsabs.harvard.edu/abs/2019MNRAS.489..241M} {489, 241}

\bibitem[\protect\citeauthoryear{{Metzger} et~al.,}{{Metzger} et~al.}{2010}]{Metzger10}
{Metzger} B.~D.,  et~al., 2010, \mn@doi [\mnras] {10.1111/j.1365-2966.2010.16864.x}, \href {http://adsabs.harvard.edu/abs/2010MNRAS.406.2650M} {406, 2650}

\bibitem[\protect\citeauthoryear{{M{\"o}ller} et~al.,}{{M{\"o}ller} et~al.}{2021}]{Moller21}
{M{\"o}ller} A.,  et~al., 2021, \mn@doi [\mnras] {10.1093/mnras/staa3602}, \href {https://ui.adsabs.harvard.edu/abs/2021MNRAS.501.3272M} {501, 3272}

\bibitem[\protect\citeauthoryear{{Morgan} et~al.,}{{Morgan} et~al.}{2020}]{Morgan20}
{Morgan} R.,  et~al., 2020, \mn@doi [\apj] {10.3847/1538-4357/abafaa}, \href {https://ui.adsabs.harvard.edu/abs/2020ApJ...901...83M} {901, 83}

\bibitem[\protect\citeauthoryear{{Muthukrishna}, {Narayan}, {Mandel}, {Biswas}  \& {Hlo{\v{z}}ek}}{{Muthukrishna} et~al.}{2019}]{Muthukrishna19}
{Muthukrishna} D.,  {Narayan} G.,  {Mandel} K.~S.,  {Biswas} R.,   {Hlo{\v{z}}ek} R.,  2019, \mn@doi [\pasp] {10.1088/1538-3873/ab1609}, \href {https://ui.adsabs.harvard.edu/abs/2019PASP..131k8002M} {131, 118002}

\bibitem[\protect\citeauthoryear{{Narayan} et~al.,}{{Narayan} et~al.}{2018}]{Narayan18}
{Narayan} G.,  et~al., 2018, \mn@doi [\apjs] {10.3847/1538-4365/aab781}, \href {https://ui.adsabs.harvard.edu/abs/2018ApJS..236....9N} {236, 9}

\bibitem[\protect\citeauthoryear{{Oke} \& {Gunn}}{{Oke} \& {Gunn}}{1983}]{Oke83}
{Oke} J.~B.,  {Gunn} J.~E.,  1983, \mn@doi [\apj] {10.1086/160817}, \href {https://ui.adsabs.harvard.edu/abs/1983ApJ...266..713O} {266, 713}

\bibitem[\protect\citeauthoryear{{Paterson} et~al.,}{{Paterson} et~al.}{2020}]{Paterson20}
{Paterson} K.,  et~al., 2020, arXiv e-prints, \href {https://ui.adsabs.harvard.edu/abs/2020arXiv201211700P} {p. arXiv:2012.11700}

\bibitem[\protect\citeauthoryear{{Perlmutter}}{{Perlmutter}}{1989}]{Perlmutter89}
{Perlmutter} S.,  1989, in Particle Astrophysics: Forefront Experimental Issues. p.~196

\bibitem[\protect\citeauthoryear{{Perlmutter} et~al.,}{{Perlmutter} et~al.}{1999}]{Perlmutter99}
{Perlmutter} S.,  et~al., 1999, \mn@doi [\apj] {10.1086/307221}, \href {https://ui.adsabs.harvard.edu/abs/1999ApJ...517..565P} {517, 565}

\bibitem[\protect\citeauthoryear{Petrov et~al.,}{Petrov et~al.}{2022}]{Petrov_2022}
Petrov P.,  et~al., 2022, \mn@doi [The Astrophysical Journal] {10.3847/1538-4357/ac366d}, 924, 54

\bibitem[\protect\citeauthoryear{{Planck Collaboration} et~al.,}{{Planck Collaboration} et~al.}{2020}]{planckcolab}
{Planck Collaboration} et~al., 2020, \mn@doi [\aap] {10.1051/0004-6361/201833910}, \href {https://ui.adsabs.harvard.edu/abs/2020A&A...641A...6P} {641, A6}

\bibitem[\protect\citeauthoryear{{Pozanenko}, {Minaev}, {Grebenev}  \& {Chelovekov}}{{Pozanenko} et~al.}{2020}]{Pozanenko20}
{Pozanenko} A.~S.,  {Minaev} P.~Y.,  {Grebenev} S.~A.,   {Chelovekov} I.~V.,  2020, \mn@doi [Astronomy Letters] {10.1134/S1063773719110057}, \href {https://ui.adsabs.harvard.edu/abs/2020AstL...45..710P} {45, 710}

\bibitem[\protect\citeauthoryear{{Richmond}, {Treffers}  \& {Filippenko}}{{Richmond} et~al.}{1993}]{Richmond93}
{Richmond} M.,  {Treffers} R.~R.,   {Filippenko} A.~V.,  1993, \mn@doi [\pasp] {10.1086/133294}, \href {https://ui.adsabs.harvard.edu/abs/1993PASP..105.1164R} {105, 1164}

\bibitem[\protect\citeauthoryear{{Riess} et~al.,}{{Riess} et~al.}{1998}]{Riess98}
{Riess} A.~G.,  et~al., 1998, \mn@doi [\aj] {10.1086/300499}, \href {https://ui.adsabs.harvard.edu/abs/1998AJ....116.1009R} {116, 1009}

\bibitem[\protect\citeauthoryear{Rodríguez-Ramírez, Bom, Fraga  \& Nemmen}{Rodríguez-Ramírez et~al.}{2023}]{rodríguezramírez2023optical}
Rodríguez-Ramírez J.~C.,  Bom C.~R.,  Fraga B.,   Nemmen R.,  2023, Optical Emission Model for Binary Black Hole Merger Remnants Travelling through Discs of Active Galactic Nucleus (\mn@eprint {arXiv} {2304.10567})

\bibitem[\protect\citeauthoryear{{Santos}, {Kilpatrick}, {Bom}, {Lacerda}  \& {STEP-GW Collaboration}}{{Santos} et~al.}{2023}]{2023GCN.33986}
{Santos} A.,  {Kilpatrick} C.~D.,  {Bom} C.~R.,  {Lacerda} E.,   {STEP-GW Collaboration} 2023, GRB Coordinates Network, \href {https://ui.adsabs.harvard.edu/abs/2023GCN.33986....1S} {33986, 1}

\bibitem[\protect\citeauthoryear{{Schechter}, {Mateo}  \& {Saha}}{{Schechter} et~al.}{1993}]{1993PASP..105.1342S}
{Schechter} P.~L.,  {Mateo} M.,   {Saha} A.,  1993, \mn@doi [\pasp] {10.1086/133316}, \href {https://ui.adsabs.harvard.edu/abs/1993PASP..105.1342S} {105, 1342}

\bibitem[\protect\citeauthoryear{{Scolnic} et~al.,}{{Scolnic} et~al.}{2015}]{Scolnic15}
{Scolnic} D.,  et~al., 2015, \mn@doi [\apj] {10.1088/0004-637X/815/2/117}, \href {http://adsabs.harvard.edu/abs/2015ApJ...815..117S} {815, 117}

\bibitem[\protect\citeauthoryear{{Shandonay} et~al.,}{{Shandonay} et~al.}{2022}]{Shandonay22}
{Shandonay} A.,  et~al., 2022, \mn@doi [\apj] {10.3847/1538-4357/ac3760}, \href {https://ui.adsabs.harvard.edu/abs/2022ApJ...925...44S} {925, 44}

\bibitem[\protect\citeauthoryear{{Simonyan} \& {Zisserman}}{{Simonyan} \& {Zisserman}}{2014}]{Simonyan14}
{Simonyan} K.,  {Zisserman} A.,  2014, \mn@doi [arXiv e-prints] {10.48550/arXiv.1409.1556}, \href {https://ui.adsabs.harvard.edu/abs/2014arXiv1409.1556S} {p. arXiv:1409.1556}

\bibitem[\protect\citeauthoryear{Singer \& Price}{Singer \& Price}{2016}]{PhysRevD.93.024013}
Singer L.~P.,  Price L.~R.,  2016, \mn@doi [Phys. Rev. D] {10.1103/PhysRevD.93.024013}, 93, 024013

\bibitem[\protect\citeauthoryear{{Smith}}{{Smith}}{2014}]{Smith14}
{Smith} N.,  2014, \mn@doi [\araa] {10.1146/annurev-astro-081913-040025}, \href {https://ui.adsabs.harvard.edu/abs/2014ARA&A..52..487S} {52, 487}

\bibitem[\protect\citeauthoryear{Soares-Santos et~al.,}{Soares-Santos et~al.}{2017}]{soares2017electromagnetic}
Soares-Santos M.,  et~al., 2017, The Astrophysical Journal Letters, 848, L16

\bibitem[\protect\citeauthoryear{Soares-Santos et~al.,}{Soares-Santos et~al.}{2019}]{Soares-Santos_2019}
Soares-Santos M.,  et~al., 2019, \mn@doi [The Astrophysical Journal Letters] {10.3847/2041-8213/ab14f1}, 876, L7

\bibitem[\protect\citeauthoryear{{Soderberg} et~al.,}{{Soderberg} et~al.}{2010}]{Soderberg10}
{Soderberg} A.~M.,  et~al., 2010, \mn@doi [\nat] {10.1038/nature08714}, \href {https://ui.adsabs.harvard.edu/abs/2010Natur.463..513S} {463, 513}

\bibitem[\protect\citeauthoryear{{Sravan}, {Milisavljevic}, {Reynolds}, {Lentner}  \& {Linvill}}{{Sravan} et~al.}{2020}]{Sravan20}
{Sravan} N.,  {Milisavljevic} D.,  {Reynolds} J.~M.,  {Lentner} G.,   {Linvill} M.,  2020, \mn@doi [\apj] {10.3847/1538-4357/ab8128}, \href {https://ui.adsabs.harvard.edu/abs/2020ApJ...893..127S} {893, 127}

\bibitem[\protect\citeauthoryear{Stein et~al.,}{Stein et~al.}{2023}]{stein_23}
Stein R.,  et~al., 2023, \mn@doi [Monthly Notices of the Royal Astronomical Society] {10.1093/mnras/stad767}, 521, 5046

\bibitem[\protect\citeauthoryear{{Szegedy}, {Vanhoucke}, {Ioffe}, {Shlens}  \& {Wojna}}{{Szegedy} et~al.}{2015}]{Szegedy2015}
{Szegedy} C.,  {Vanhoucke} V.,  {Ioffe} S.,  {Shlens} J.,   {Wojna} Z.,  2015, CoRR, abs/1512.00567

\bibitem[\protect\citeauthoryear{{Tagawa}, {Kimura}, {Haiman}, {Perna}  \& {Bartos}}{{Tagawa} et~al.}{2023}]{tagawa23}
{Tagawa} H.,  {Kimura} S.~S.,  {Haiman} Z.,  {Perna} R.,   {Bartos} I.,  2023, \mn@doi [arXiv e-prints] {10.48550/arXiv.2303.02172}, \href {https://ui.adsabs.harvard.edu/abs/2023arXiv230302172T} {p. arXiv:2303.02172}

\bibitem[\protect\citeauthoryear{{Thakur} et~al.,}{{Thakur} et~al.}{2020}]{Thakur20}
{Thakur} A.~L.,  et~al., 2020, \mn@doi [\mnras] {10.1093/mnras/staa2798}, \href {https://ui.adsabs.harvard.edu/abs/2020MNRAS.499.3868T} {499, 3868}

\bibitem[\protect\citeauthoryear{{Tody}}{{Tody}}{1986}]{IRAF}
{Tody} D.,  1986, in {Crawford} D.~L.,  ed.,  Society of Photo-Optical Instrumentation Engineers (SPIE) Conference Series Vol. 627, Instrumentation in astronomy VI. p.~733, \mn@doi{10.1117/12.968154}

\bibitem[\protect\citeauthoryear{{Tonry} et~al.,}{{Tonry} et~al.}{2022}]{2022tiv-disc}
{Tonry} J.,  et~al., 2022, Transient Name Server Discovery Report, \href {https://ui.adsabs.harvard.edu/abs/2022TNSTR2574....1T} {2022-2574, 1}

\bibitem[\protect\citeauthoryear{{Toy}, {Grayling}, {Wiseman}, {Frohmaier}  \& {Yaron}}{{Toy} et~al.}{2022}]{2022TNSCR2660....1T}
{Toy} M.,  {Grayling} M.,  {Wiseman} P.,  {Frohmaier} C.,   {Yaron} O.,  2022, Transient Name Server Classification Report, \href {https://ui.adsabs.harvard.edu/abs/2022TNSCR2660....1T} {2022-2660, 1}

\bibitem[\protect\citeauthoryear{{Tucker} et~al.,}{{Tucker} et~al.}{2022}]{Tucker21}
{Tucker} D.~L.,  et~al., 2022, \mn@doi [\apj] {10.3847/1538-4357/ac5b60}, \href {https://ui.adsabs.harvard.edu/abs/2022ApJ...929..115T} {929, 115}

\bibitem[\protect\citeauthoryear{{Vieira} et~al.,}{{Vieira} et~al.}{2020}]{Vieira20}
{Vieira} N.,  et~al., 2020, \mn@doi [\apj] {10.3847/1538-4357/ab917d}, \href {https://ui.adsabs.harvard.edu/abs/2020ApJ...895...96V} {895, 96}

\bibitem[\protect\citeauthoryear{{Villar} et~al.,}{{Villar} et~al.}{2019}]{Villar19}
{Villar} V.~A.,  et~al., 2019, \mn@doi [\apj] {10.3847/1538-4357/ab418c}, \href {https://ui.adsabs.harvard.edu/abs/2019ApJ...884...83V} {884, 83}

\bibitem[\protect\citeauthoryear{{Volgenau}, {Harbeck}, {Lindstrom}, {Collom}, {Street}  \& {Johnson}}{{Volgenau} et~al.}{2022}]{Volgenau22}
{Volgenau} N.,  {Harbeck} D.,  {Lindstrom} W.,  {Collom} D.,  {Street} R.,   {Johnson} M.,  2022, in {Adler} D.~S.,  {Seaman} R.~L.,   {Benn} C.~R.,  eds,  Society of Photo-Optical Instrumentation Engineers (SPIE) Conference Series Vol. 12186, Observatory Operations: Strategies, Processes, and Systems IX. p. 121860W, \mn@doi{10.1117/12.2628704}

\bibitem[\protect\citeauthoryear{{Watson} et~al.,}{{Watson} et~al.}{2020}]{Watson20}
{Watson} A.~M.,  et~al., 2020, \mn@doi [\mnras] {10.1093/mnras/staa161}, \href {https://ui.adsabs.harvard.edu/abs/2020MNRAS.492.5916W} {492, 5916}

\bibitem[\protect\citeauthoryear{{Wells}, {Greisen}  \& {Harten}}{{Wells} et~al.}{1981}]{FITS}
{Wells} D.~C.,  {Greisen} E.~W.,   {Harten} R.~H.,  1981, \aaps, \href {https://ui.adsabs.harvard.edu/abs/1981A&AS...44..363W} {44, 363}

\bibitem[\protect\citeauthoryear{{Wolf} et~al.,}{{Wolf} et~al.}{2018}]{skymapperDR1}
{Wolf} C.,  et~al., 2018, \mn@doi [\pasa] {10.1017/pasa.2018.5}, \href {https://ui.adsabs.harvard.edu/abs/2018PASA...35...10W} {35, e010}

\bibitem[\protect\citeauthoryear{{Woosley}}{{Woosley}}{1988}]{Woosley88}
{Woosley} S.~E.,  1988, \mn@doi [\apj] {10.1086/166468}, \href {https://ui.adsabs.harvard.edu/abs/1988ApJ...330..218W} {330, 218}

\bibitem[\protect\citeauthoryear{{Wyatt}, {Tohuvavohu}, {Arcavi}, {Lundquist}, {Howell}  \& {Sand}}{{Wyatt} et~al.}{2020}]{Wyatt20}
{Wyatt} S.~D.,  {Tohuvavohu} A.,  {Arcavi} I.,  {Lundquist} M.~J.,  {Howell} D.~A.,   {Sand} D.~J.,  2020, \mn@doi [\apj] {10.3847/1538-4357/ab855e}, \href {https://ui.adsabs.harvard.edu/abs/2020ApJ...894..127W} {894, 127}

\bibitem[\protect\citeauthoryear{{Yang} et~al.,}{{Yang} et~al.}{2023}]{yang_kn23}
{Yang} Y.-H.,  et~al., 2023, \mn@doi [arXiv e-prints] {10.48550/arXiv.2308.00638}, \href {https://ui.adsabs.harvard.edu/abs/2023arXiv230800638Y} {p. arXiv:2308.00638}

\bibitem[\protect\citeauthoryear{{Yaron} \& {Gal-Yam}}{{Yaron} \& {Gal-Yam}}{2012}]{Yaron12}
{Yaron} O.,  {Gal-Yam} A.,  2012, \mn@doi [\pasp] {10.1086/666656}, \href {https://ui.adsabs.harvard.edu/abs/2012PASP..124..668Y} {124, 668}

\bibitem[\protect\citeauthoryear{{Zhou} et~al.,}{{Zhou} et~al.}{2021}]{Zhou20}
{Zhou} R.,  et~al., 2021, \mn@doi [\mnras] {10.1093/mnras/staa3764}, \href {https://ui.adsabs.harvard.edu/abs/2021MNRAS.501.3309Z} {501, 3309}

\bibitem[\protect\citeauthoryear{Zwicky}{Zwicky}{1938}]{zwicky1938search}
Zwicky F.,  1938, Publications of the Astronomical Society of the Pacific, 50, 215

\bibitem[\protect\citeauthoryear{{de Wet} et~al.,}{{de Wet} et~al.}{2021}]{deWet21}
{de Wet} S.,  et~al., 2021, arXiv e-prints, \href {https://ui.adsabs.harvard.edu/abs/2021arXiv210302399D} {p. arXiv:2103.02399}

\makeatother
\end{thebibliography}

\appendix
\appendix
\section{Photometry of Followed Targets}\label{appendix:photometry}

In this appendix, we present the photometry of targets, followed by STEP in its first year of operations. All photometry is obtained in difference imaging on the AB magnitude system \citep{Oke83} and does not consider dust extinction and other colour correction terms.

\begin{table}
\centering
    \begin{tabular}{cccc}
    \hline
    \multicolumn{4}{c}{2022crv} \\ \hline
    MJD & Band & $m$ & $\sigma_{m}$ \\ 
    & & (mag) & (mag) \\ \hline\hline
    59648.0185 & $g$ & 15.799 & 0.015 \\
    59650.0208 & $g$ & 15.892 & 0.016 \\
    59653.0853 & $g$ & 16.159 & 0.021 \\
    59655.1146 & $g$ & 16.319 & 0.025 \\
    59663.0323 & $g$ & 16.992 & 0.018 \\
    59665.0080 & $g$ & 17.119 & 0.020 \\
    59667.0067 & $g$ & 17.146 & 0.020 \\
    59669.0099 & $g$ & 17.233 & 0.019 \\
    59673.0019 & $g$ & 17.396 & 0.020 \\
    59674.9997 & $g$ & 17.497 & 0.022 \\
    59648.0240 & $i$ & 15.384 & 0.009 \\
    59650.0270 & $i$ & 15.367 & 0.009 \\
    59653.0909 & $i$ & 15.402 & 0.010 \\
    59655.1201 & $i$ & 15.472 & 0.010 \\
    59657.0362 & $i$ & 15.586 & 0.012 \\
    59659.1176 & $i$ & 15.674 & 0.012 \\
    59661.0167 & $i$ & 15.759 & 0.010 \\
    59663.0379 & $i$ & 15.859 & 0.008 \\
    59665.0136 & $i$ & 15.940 & 0.009 \\
    59667.0119 & $i$ & 16.042 & 0.009 \\
    59669.0160 & $i$ & 16.111 & 0.009 \\
    59671.0086 & $i$ & 16.202 & 0.010 \\
    59673.0067 & $i$ & 16.239 & 0.010 \\
    59675.0053 & $i$ & 16.325 & 0.010 \\
    59677.0040 & $i$ & 16.407 & 0.010 \\
    59679.0025 & $i$ & 16.456 & 0.012 \\
    59681.0010 & $i$ & 16.499 & 0.012 \\
    59683.0279 & $i$ & 16.515 & 0.014 \\
    59648.0212 & $r$ & 15.336 & 0.008 \\
    59650.0243 & $r$ & 15.370 & 0.010 \\
    59653.0881 & $r$ & 15.471 & 0.010 \\
    59655.1173 & $r$ & 15.585 & 0.010 \\
    59657.0334 & $r$ & 15.760 & 0.013 \\
    59659.1148 & $r$ & 15.852 & 0.013 \\
    59661.0139 & $r$ & 15.982 & 0.011 \\
    59663.0350 & $r$ & 16.090 & 0.009 \\
    59665.0107 & $r$ & 16.201 & 0.010 \\
    59667.0090 & $r$ & 16.302 & 0.009 \\
    59669.0131 & $r$ & 16.356 & 0.011 \\
    59671.0055 & $r$ & 16.469 & 0.012 \\
    59673.0038 & $r$ & 16.531 & 0.010 \\
    59675.0024 & $r$ & 16.608 & 0.012 \\
    59677.0011 & $r$ & 16.669 & 0.013 \\
    59678.9996 & $r$ & 16.697 & 0.014 \\
    59680.9981 & $r$ & 16.759 & 0.015 \\
    59683.0240 & $r$ & 16.830 & 0.017 \\
    59648.0269 & $z$ & 15.351 & 0.011 \\
    59650.0300 & $z$ & 15.372 & 0.012 \\
    59653.0938 & $z$ & 15.380 & 0.012 \\
    59655.1230 & $z$ & 15.427 & 0.011 \\
    59657.0392 & $z$ & 15.525 & 0.012 \\
    59659.1206 & $z$ & 15.584 & 0.013 \\
    59661.0197 & $z$ & 15.681 & 0.013 \\
    59663.0412 & $z$ & 15.747 & 0.011 \\
    59665.0168 & $z$ & 15.833 & 0.011 \\
    59667.0152 & $z$ & 15.895 & 0.009 \\
    59669.0193 & $z$ & 15.921 & 0.009 \\
    59671.0118 & $z$ & 16.011 & 0.012 \\
    59673.0100 & $z$ & 16.037 & 0.011 \\
    59675.0085 & $z$ & 16.107 & 0.012 \\
    59677.0073 & $z$ & 16.153 & 0.011 \\
    59679.0057 & $z$ & 16.177 & 0.014 \\
    59681.0037 & $z$ & 16.220 & 0.013 \\
    59683.0318 & $z$ & 16.231 & 0.014 \\
    \hline
    \end{tabular}
    \caption{Light curve data for SN 2022crv.}
\end{table}

\begin{table}
\centering
    \begin{tabular}{cccc}
    \multicolumn{4}{c}{2022acko}\\ \hline
    MJD & Band & $m$ & $\sigma_{m}$ \\ 
    & & (mag) & (mag) \\ \hline\hline
    59933.0474 & $g$ & 16.215 & 0.042 \\
    59933.0487 & $r$ & 16.051 & 0.039 \\
    59933.0501 & $i$ & 16.241 & 0.052 \\
    59933.0514 & $z$ & 16.394 & 0.075 \\
    59939.1292 & $g$ & 16.509 & 0.056 \\
    59939.1305 & $r$ & 16.111 & 0.041 \\
    59939.1318 & $i$ & 16.210 & 0.059 \\
    59939.1332 & $z$ & 16.265 & 0.080 \\
    59943.1096 & $r$ & 16.139 & 0.049 \\
    59943.1111 & $i$ & 16.261 & 0.061 \\
    59943.1123 & $z$ & 16.221 & 0.078 \\
    \hline
    \end{tabular}
    \caption{Light curve data for SN 2022acko.}
\end{table}

\begin{table}
\centering
    \begin{tabular}{cccc}
    \hline
    \multicolumn{4}{c}{2022qxu}\\ \hline
    MJD & Band & $m$ & $\sigma_{m}$ \\ 
    & & (mag) & (mag) \\ \hline\hline
    59815.2268 & $g$ & >16.221 & -- \\
    59819.3962 & $g$ & >16.221 & -- \\
    59824.3020 & $g$ & >16.221 & -- \\
    59828.2908 & $g$ & >16.221 & -- \\
    59837.3453 & $g$ & >16.221 & -- \\
    59845.3483 & $g$ & >16.221 & -- \\
    59851.2632 & $g$ & >16.221 & -- \\
    59815.2325 & $i$ & 20.489 & 0.194 \\
    59819.4024 & $i$ & 20.562 & 0.254 \\
    59824.3080 & $i$ & 20.682 & 0.292 \\
    59828.2966 & $i$ & 20.822 & 0.320 \\
    59837.3511 & $i$ & >20.822 & -- \\
    59841.1792 & $i$ & >20.822 & -- \\
    59845.3541 & $i$ & >20.822 & -- \\
    59851.2690 & $i$ & >20.822 & -- \\
    59815.2293 & $r$ & 20.698 & 0.198 \\
    59819.3995 & $r$ & 20.789 & 0.146 \\
    59824.3046 & $r$ & 20.945 & 0.211 \\
    59828.2936 & $r$ & 21.241 & 0.330 \\
    59837.3481 & $r$ & >21.241 & -- \\
    59845.3511 & $r$ & >21.241 & -- \\
    59851.2660 & $r$ & >21.241 & -- \\
    59815.2361 & $z$ & 20.498 & 0.193 \\
    59819.4062 & $z$ & 20.512 & 0.210 \\
    59824.3119 & $z$ & 20.652 & 0.221 \\
    59828.3000 & $z$ & 20.710 & 0.330 \\
    59837.3544 & $z$ & >20.710 & -- \\
    59841.1825 & $z$ & >20.710 & -- \\
    59845.3574 & $z$ & >20.710 & -- \\
    59851.2723 & $z$ & >20.710 & -- \\
    \end{tabular}
    \caption{Light curve data for SN 2022qxu.}
\end{table}

\bsp	
\label{lastpage}
\end{document}